\documentclass[11pt]{article}
\pdfoutput=1
\usepackage{jheppub}
\usepackage{tikz}
\usetikzlibrary{calc}
\usepackage{graphicx,subcaption}
 \usepackage{mathtools}

\newcommand\fft[2]{{\frac{#1}{#2}}}

\renewcommand{\Re}{\operatorname{Re}}
\renewcommand{\Im}{\operatorname{Im}}
\newcommand{\Tr}{\operatorname{Tr}}

\newcommand{\CA}{\mathcal{A}}

\newcommand{\CM}{\mathcal{M}}
\newcommand{\CN}{\mathcal{N}}

\newcommand{\CD}{\mathcal{D}}
\newcommand{\CH}{\mathcal{H}}

\newcommand{\CI}{\mathcal I}
\newcommand{\CQ}{\mathcal Q}

\newcommand{\sgn}{\mbox{sgn}}

\newcommand{\IR}{\mathbb{R}}
\newcommand{\IC}{\mathbb{C}}
\newcommand{\IZ}{\mathbb{Z}}

\renewcommand{\=}{\;= \;}
\newcommand{\+}{\;+ \;}

\newcommand{\nn}{\nonumber}
\newcommand{\dd}{\mathrm{d}}

\newcommand{\rme}{{\rm e}}

\newcommand\be{\begin{equation}}
\newcommand\ee{\end{equation}}
\newcommand\bea{\begin{eqnarray}}
\newcommand\eea{\end{eqnarray}}

\renewcommand{\O}{\Omega}

\renewcommand{\a}{\alpha}
\renewcommand{\b}{\beta}
\newcommand{\s}{\sigma}
\renewcommand{\t}{\tau}
\newcommand{\g}{\gamma}

\newcommand{\wt}{\widetilde}

\renewcommand{\Re}{\text{Re}}
\renewcommand{\Im}{\text{Im}}

\newcommand{\ve}{\varepsilon}

\newcommand{\ndt}{\noindent}

\renewcommand{\i}{{\rm i}}

\newcommand{\Ge}{\Gamma_\text{e}}

\newcommand{\z}{\zeta}

\newcommand{\defeq}{\; \coloneqq \;} 

\newcommand{\CQb}{\overline{\mathcal{Q}}}
\newcommand{\Seff}{S_\text{eff}}

\newcommand{\rO}{\text{O}}

\newcommand{\uu}{\underline{u}}
\newcommand{\ux}{\underline{x}}
\newcommand{\um}{\underline{m}}

\newcommand{\rr}{\gamma}

\newcommand{\tf}{\text{f}}

\begin{document}

\title{The 4d superconformal index near roots of unity and \\ 3d Chern-Simons theory}

\author{Arash Arabi Ardehali and Sameer Murthy}
\affiliation{Department of Mathematics, King's College London,\\
The Strand, London WC2R 2LS, U.K.}

\emailAdd{a.a.ardehali@gmail.com}
\emailAdd{sameer.murthy@kcl.ac.uk}

\abstract{ 
We consider the~$S^3\times S^1$ superconformal index~$\mathcal{I}(\tau)$ of 4d~$\CN=1$ 
gauge theories. The Hamiltonian index is defined in a standard manner as  
the Witten index with a chemical potential~$\tau$ coupled to a combination of angular 
momenta on~$S^3$ and the~$U(1)$ R-charge. We develop the all-order asymptotic 
expansion of the index as~$q = e^{2 \pi {\rm i} \tau}$ approaches a root of unity, 
i.e.~as~$\widetilde \tau \equiv m \tau + n \to 0$, with $m,n$ relatively prime integers. 
The asymptotic expansion of $\log\mathcal{I}(\tau)$ has terms of the 
form~$\widetilde \tau^k$, $k = -2, -1, 0, 1$. We determine the coefficients of the~$k=-2,-1,1$ 
terms from the gauge theory data, and provide evidence that 
the~$k=0$ term is determined by the Chern-Simons partition function on~$S^3/\mathbb{Z}_m$. 
We explain these findings from the point of view of the 3d theory obtained by reducing the 
4d gauge theory on the~$S^1$. The supersymmetric functional integral of the 3d theory 
takes the form of a matrix integral over the dynamical 3d fields, with an effective action 
given by supersymmetrized Chern-Simons couplings of background and dynamical 
gauge fields. The singular terms in the~$\widetilde \tau \to 0$ expansion (dictating 
the growth of the 4d index) are governed by the background 
Chern-Simons couplings. The constant term has a background piece as well as a piece given by the localized functional 
integral over the dynamical 3d gauge multiplet. The linear term arises from the supersymmetric 
Casimir energy factor needed to go between the functional integral and the Hamiltonian index.
}

\maketitle \flushbottom

\section{Introduction}

In the last few years there has been a renewed interest in 
the study of the superconformal index of 4d~$\CN=1$ superconformal field theories (SCFTs) and,
in particular, $\CN=4$ super Yang-Mills (SYM). 
The index in question is the supersymmetric partition function of the SCFT on~$S^3 \times S^1$
which receives contributions from BPS states that preserve two supercharges~$(\CQ, \CQb)$. 
In the large-$N$ limit, the expectation from AdS/CFT 
is that the index should account for the entropy of the BPS black holes (BH) 
that preserve the same two supercharges in the dual supergravity on~AdS$_5$. 
This question was introduced in~\cite{Sundborg:1999ue, Aharony:2003sx, Kinney:2005ej}, 
and the work of the last few years has shown that the index indeed captures the BH entropy
in different asymptotic 
limits~\cite{Hosseini:2017mds,Cabo-Bizet:2018ehj, Choi:2018hmj,Benini:2018ywd,
Choi:2018vbz, Honda:2019cio,ArabiArdehali:2019tdm,Kim:2019yrz,Cabo-Bizet:2019osg,
Cabo-Bizet:2019eaf, Benini:2020gjh, Amariti:2019mgp,Lezcano:2019pae,Lanir:2019abx,David:2020ems, 
Cabo-Bizet:2020nkr,Murthy:2020rbd,Agarwal:2020zwm,Cabo-Bizet:2020ewf,Copetti:2020dil,Goldstein:2020yvj}. 

The focus of the present paper is the \emph{Cardy-like limit} in which the BH entropy 
becomes very large. In the canonical ensemble, this translates to the study of the exponential growth of the 
index as~$\t \to 0$, where the parameter~$\t$ is the chemical potential dual to the charge.
As pointed out in~\cite{Cabo-Bizet:2019eaf}, the $\t \to 0$ limit is in fact one of an infinite number of 
inequivalent Cardy-like limits in which the index is expected to grow exponentially. 
These limits correspond to~$\t$ approaching a rational number or, equivalently,~$q = e^{2\pi \i \t}$ 
approaching a root of unity. 
In this paper we analyze the 4d superconformal index near a general root of unity, 
and find interesting relations to three-dimensional Chern-Simons (CS) theory.
The main statement is that  the asymptotics of the index near a rational point~$-n/m$ is equal 
(to all orders in perturbation theory in deviations~$\wt \t = m\t +n$ from the rational point) to 
the partition function of a certain 3d $\mathcal{N}=2$ gauge theory with 
Chern-Simons couplings that involve background as well as dynamical fields on 
an~$S^3/\mathbb{Z}_m$ orbifold. The background couplings give rise to singular terms 
at~$\rO(1/\wt \t^2)$ and~$\rO(1/\wt \t)$ that govern the growth of the index, while the 
constant~$\rO(1)$ term receives contributions from both background fields and 
the dynamical Chern-Simons theory.  

We demonstrate this statement from two points of view---by direct asymptotic analysis of 
the index near rational points, and from an analysis of the reduced three-dimensional theory and calculating the
various couplings using high-temperature effective-field theory (EFT) techniques. 
The latter method, based on~\cite{DiPietro:2014bca, DiPietro:2016ond}, 
relates the high-temperature asymptotics 
of the index to a low-energy effective 
field theory, in the spirit of the Cardy formula.\footnote{In the high-temperature picture 
the (Euclidean) time direction is taken along the $S^1$, while in the low-energy picture 
time is a fiber inside the $S^3$. Relating the two pictures involves swapping time and 
space as in the derivation of the 2d Cardy formula~\cite{Cardy:1986ie}. Unlike in the 
two-dimensional context where one uses~$SL(2,\IZ)$ automorphy to relate the 
swapped problem to the original one, here we do not have an a priori understanding 
of the automorphic properties of the 4d index~$\CI(\tau)$. Aspects of this question are being 
addressed in~\cite{GMZ}. See also~\cite{Shaghoulian:2016gol,Gadde:2020bov} for 
related work on modular-type transformation properties relating different indices, 
and~\cite{Razamat:2012uv} for a discussion of the automorphic behavior of a different 
index in $\mathcal{N}=2$ SCFTs.}

\vskip 0.4cm

\ndt {\bf The four-dimensional superconformal index and its asymptotic growth}

\vskip 0.1cm

In this paper we study~$\CN=1$ gauge theories with a Lagrangian description and a $U(1)_R$ symmetry, 
with a focus on~$\CN=4$ SYM which we use to illustrate some statements in detail. 
The symmetry algebra of $\mathcal{N}=1$ SCFT on $S^1\times S^3$ is~$SU(2,2|1)$,
which includes the energy~$E$ which generates translations around~$S^1$, 
the angular momenta~$J_1$, $J_2$ on~$S^3$, 
and the $U(1)$ R-charge~$Q$. 
One can pick a complex supercharge obeying the following algebra, 
\be \label{QQbaralg}
\{\CQ, \CQb \} \= E-J_1-J_2-\tfrac{3}{2}\,Q \,.
\ee
The most general index built out of the~$\CN=1$ superconformal algebra is an extension of the 
Witten index of~$\CQ$ and is defined as the following trace over the physical Hilbert space,
\be \label{defindex}
\CI(\s, \t) 
\=  \, {\rm Tr}_{\CH}\,  (-1)^F \rme^{- \g \{\CQ, \CQb \}   
+2 \pi \i \s (J_1+\frac{1}{2}Q)+2 \pi \i \t (J_2+\frac{1}{2}Q)} \,.
\ee
The trace~\eqref{defindex} only receives contributions from states annihilated by the supercharges 
($\frac14$-BPS states) so that the right-hand side of~\eqref{QQbaralg} vanishes for these states.  
This index~$\CI(\s,\t)$ can be calculated from either Hamiltonian or functional integral methods 
and reduces to a unitary matrix integral~\cite{Romelsberger:2005eg,Kinney:2005ej,Nawata:2011un,Assel:2014paa}, 
which can be written as an  integral over the space of gauge holonomies around the~$S^1$
of certain infinite products, as written in Equation~\eqref{eq:pqIndex}.

Our focus in this paper is the analog, in the present context, of the 
high-temperature Cardy limit of 2d CFT. This means fixing the rank and taking the 
charges~$(J_i,Q)$ to be larger than any other scale in the theory.
In the canonical ensemble this translates to taking~$\Im \, \s, \, \Im \, \t \to 0$ at fixed rank. 
In order to calculate the asymptotic growth of states along a certain direction in the charge lattice, 
one needs to fix the relation between~$\s$ and~$\t$.
We study\footnote{Our methods can be generalized to study the case where~$\s$ and~$\t$ are linearly
dependent over the rationals, but we shall not develop this in the present paper.
} 
the slice~$\s=\t-n_0$ with~$n_0$ an integer, 
as in~\cite{Cabo-Bizet:2019osg, Cabo-Bizet:2019eaf, Cabo-Bizet:2020nkr}.
Setting~$2J =J_1+J_2$, the resulting canonical index~$\CI$ is given by
\be \label{eq:n0Index}
\CI(\t;n_0) \=  \, {\rm Tr}_{\CH}\,  (-1)^F \rme^{-\g \{\CQ, \CQb \}   
-2 \pi \i n_0 (J_1+\frac{1}{2}Q)+2 \pi \i \t (2J+Q)} \,.
\ee
The large-charge asymptotics then implies~$\Im\,\t \to 0$, 
while~$\Re\,\t$ is not fixed a priori by the limit. 
We consider asymptotic limits as~$\t$ approaches a rational 
number~$\t \to -n/m$ with~$\text{gcd}(m,n)=1$,
introduced in the present context as new Cardy-like limits in~\cite{Cabo-Bizet:2019eaf}. 
The index $\mathcal{I}$ clearly depends on the gauge group $G$. We generally suppress 
it in our notation, but sometimes use the notation $\mathcal{I}_N$ to emphasize the 
dependence on $N$ for $U(N)$ or $SU(N)$ $\mathcal{N}=4$ SYM theory (which 
should be clear from the context). 

Our motivation to consider these rational points comes from the study
of the index~$\CI_N(\t)$ of $\mathcal{N}=4$ SYM in the large-$N$ 
limit.\footnote{Another motivation comes from the mathematical literature on $q$-series, where 
it is also natural to consider expansions around roots of unity. We thank D.~Zagier for emphasizing this point to us.} 
In this limit one considers charges scaling as~$N^2$ as~$N \to \infty$, 
which translates to~$N \to \infty$ at fixed~$\t$ in the canonical ensemble~\cite{Murthy:2020rbd}. 
In this large-$N$ limit one expects the field theory index~$\CI_N(\t)$ to be written as a 
sum over saddles. This picture has been partially realized in the last few years using two different approaches---the 
Bethe-ansatz-like approach developed in~\cite{Closset:2017bse,Benini:2018mlo, Benini:2018ywd}, 
and the direct study of large-$N$ saddle points using an elliptic extension of the 
action~\cite{Cabo-Bizet:2019eaf, Cabo-Bizet:2020nkr}. 
In particular, the large-$N$ approach in~\cite{Cabo-Bizet:2019eaf} found a class of  
saddles labelled by rational numbers~$-n/m$, where the perturbation expansion 
around each saddle is given by the asymptotic limit~$\t \to -n/m$.\footnote{These saddles 
map to residues of the Bethe-ansatz type approach---see~\cite{Cabo-Bizet:2020ewf} 
for a recent discussion of the connections between the two approaches. 
A larger set of saddles have been classified in~\cite{Cabo-Bizet:2020nkr}, but 
the full set of important/contributing saddles is not understood in either approach.  
In particular, interesting continuum configurations of the Bethe-ansatz equations have been 
recently discovered in \cite{ArabiArdehali:2019orz,Benini:2021ano,Lezcano:2021qbj} 
whose role in the large-$N$ limit is not fully understood.}
Setting~$n_0=-1$, we have 
\be \label{mnasymp}
\log \CI_N(\t) \; \sim \; -\Seff(m,n; \tau) \,, \qquad \t \to -n/m \,, \\
\ee
where the effective action at each saddle is given by
\be\label{actionEll}
\Seff(m,n; \tau) \= \frac{N^2 \pi \i }{27\,m} \,\frac{  \bigl(2 \wt \t  +  \chi_1(m+n) \bigr)^3}{{\wt \t}^2}  \,, 
\qquad \wt \t \defeq m\t +n \,.
\ee
where~$\chi_1(n)$ is the Dirichlet character equal to~$0, \pm 1$ when~$n \equiv 0, \pm 1$ (mod~3), respectively.
There was one caveat in the above result, which was stressed in~\cite{Cabo-Bizet:2019eaf, Cabo-Bizet:2020nkr}, 
namely that the pure-imaginary~$\wt \t$-independent term could not be fixed by the methods used in those papers.
The constant term in the effective action~\eqref{mnasymp}, therefore, was a convenient choice made using 
inputs coming from outside the field-theory analysis.  

Although we do not have a rigorous notion of the sum over saddles yet,
it should be clear that if the effective action of the~$(m,n)$ saddle has negative real part it dominates 
over the others as~$m\tau +n \to 0$. 
It is also clear from~\eqref{actionEll} that the fastest growth among these saddles comes from~$(m,n)=(1,0)$. 
The~$(1,0)$ saddle 
in the SYM theory is identified as a fully deconfined phase whose 
entropy scales as~$N^2$, while the other $(m,n)$ saddles have entropies that are suppressed by a factor of~$m$.
For this reason they can be called partially deconfined saddles (in the sense of asymptotic growth, 
but not in the sense of center symmetry breaking---cf.~\cite{ArabiArdehali:2019orz}). 
On the gravitational side, the action~$\Seff(1,0; \tau)$ agrees precisely with the canonical on-shell action of the 
black hole solution in the dual AdS$_5$  supergravity~\cite{Cabo-Bizet:2018ehj},
which leads to the identification of the AdS$_5$ BH as the saddle~$(1,0)$.
The~$(m,n)$ solutions have been identified with orbifolds of the Euclidean AdS$_5$ BH~\cite{AhaSBTalk}.

Because of the dominance of the~$(1,0)$ saddle near~$\t \to 0$, one can capture it directly
in an asymptotic expansion---even for finite~$N$.
In this calculation, one writes the index~\eqref{eq:n0Index} as an integral over gauge 
holonomies~$u_i$ (see~\eqref{eq:pqIndex} below), estimates the integrand in the 
Cardy-like limit~$\t \to 0$, and then performs the integrals. 
The initial studies~\cite{Choi:2018vbz, ArabiArdehali:2019tdm,Honda:2019cio,Kim:2019yrz,Cabo-Bizet:2019osg} 
successfully reproduced the singular parts of the action as~$\t \to 0$, 
i.e.~the~$1/\t^2$ and the~$1/\t$ terms with the correct coefficients.
More recently, the complete action~\eqref{actionEll} for~$(m,n) = (1,0)$ was obtained 
in~\cite{GonzalezLezcano:2020yeb} by a direct method,  
involving a careful analysis of all perturbative terms in the Cardy-like limit. 
(See \cite{Amariti:2020jyx,Amariti:2021ubd} for more recent related work.)

Our first goal in this paper is to obtain the complete perturbative action at all the~$(m,n)$ saddles by a 
direct asymptotic analysis of the index as~$\t \to -n/m$. 
This analysis is described in Section~\ref{sec:SCI}, the result of which is a 
perfect agreement with the action~\eqref{actionEll}, up to the constant terms as mentioned above.
The asymptotic analysis requires developing the asymptotics of the elliptic gamma function 
\cite{Ruijsenaars:1997,felder2000elliptic} near rational points. The~$\tau\to0$ asymptotic 
estimates were available in previous literature~\cite{Rains:2006dfy}. Here we develop the 
analysis for~$\tau$ approaching rational numbers. The analysis is presented in 
Appendix~\ref{app:Estimates}. (See also~\cite{Kels:2017vbc} for related work motivated 
by integrable-systems considerations.)

Furthermore, we note that for given~$m,n$, depending on the sign of~$\mathrm{arg}\wt\t-\pi/2$ 
the action in~\eqref{actionEll} can have negative or positive real part, which respectively yields 
a growing or decaying contribution to the index. Therefore in essentially half of the parameter 
space the saddles in~\eqref{actionEll} do not capture any growth in the index. As demonstrated 
in Section~\ref{sec:Ccenter}, when the $(m,n)$ saddle in~\eqref{actionEll} gives a decaying 
contribution to the index as~$\wt\t\to0$, a ``2-center saddle'' takes over which yields exponential 
growth again. In other words, in half of the parameter space the growth of the 
index~$\mathcal{I}_N(\tau)$ is captured by 2-center saddles. (These turn out to be partially 
deconfined saddles both in the sense of asymptotic growth and in the sense of center 
symmetry breaking---cf.~\cite{ArabiArdehali:2019orz}.)

\vskip 0.4cm

\ndt {\bf Chern-Simons theory from the asymptotics of the 4d index}

\vskip 0.1cm

The second goal of the paper is partly inspired by an interesting pattern 
appearing in the asymptotic calculations. 
As emphasized in the context of SU($N$)~$\mathcal{N}=4$ SYM in~\cite{GonzalezLezcano:2020yeb}, 
in the part of the parameter space where the index is dominated by isolated, 1-center saddles, 
the complete asymptotic expansion in~$\t$
terminates at~$O(\t)$---i.e.~the perturbation theory only contains~$1/\t^2$, $1/\t$, $\t^0$ and~$\t$ 
up to exponentially suppressed corrections. (This is, in fact, more generally true when the index is 
dominated by isolated saddles, and not true when there are flat directions; see~\cite{Ardehali:2015bla}.) 
Interestingly, it was found in~\cite{GonzalezLezcano:2020yeb} that the constant term in the 
expansion contains the partition function of SU($N$) pure Chern-Simons theory on~$S^3$ at level~$\pm N$.

In this paper we find that the same structure persists at all rational points. We see that 
the constant term in the expansion as~$\t \to -n/m$ involves Chern-Simons theory 
whose action is~$1/m$ times the action as~$\t \to 0$. We present evidence that this corresponds 
to CS theory on an orbifold space~$S^3/\IZ_m$ (with the action of~$\IZ_m$ depending on~$n$ 
such that the orbifold coincides with the lens space~$L(m,-1)$ when~$n=1$) at 
level~$\pm N$~\cite{Garoufalidis:2006ew,Gang:2019juz}. In other words, the~4d SYM index appears to play the 
role of a master index which governs the partition function of three-dimensional CS theory on an infinite family of~$S^3$ orbifolds.

The appearance of 3d Chern-Simons theory from the 4d superconformal index is intriguing, and gives rise to 
two related questions:\\
(a) is there a direct three-dimensional physics explanation of the appearance of Chern-Simons theory?\\
(b) can we also understand the singular terms in the asymptotic expansions around 
rational points as being related to 3d Chern-Simons theory?\\
The answers to both these questions are positive, as we now explain.

\vskip 0.4cm

\ndt {\bf The asymptotics of the 4d index from supersymmetric Chern-Simons theory}

\vskip 0.1cm

The natural idea is that the reduction of the four-dimensional theory on~$S^1$ gives rise to a three-dimensional 
theory on~$S^3$ in a ``high-temperature" expansion in powers of the circumference~$\beta$ of the shrinking 
circle. 
If we calculate the functional integral of the three-dimensional theory, we should 
recover the four-dimensional functional integral as~$\beta \to 0$.   
The three-dimensional effective field theory
is known to have a derivative expansion, where the most relevant terms are  
Chern-Simons terms~\cite{Banerjee:2012iz,Jensen:2012jh}.
This EFT approach was developed in the supersymmetric context 
in~\cite{DiPietro:2014bca, DiPietro:2016ond} who presented \emph{supersymmetrized} CS actions 
involving the dynamical as well as background fields, which are 
necessary for preserving supersymmetry on~$S^3\times S^1$. In particular, 
the~$1/\beta^2$ and~$1/\beta$ effective actions derived this way in~\cite{DiPietro:2016ond} 
reproduced the asymptotics of the the index as found in~\cite{Ardehali:2015bla} for $n_0=0$ and $\mathrm{arg}(\t)=\pi/2$.
(Note that when the metric on $S^3\times S^1$ has a direct product form with $S^3$ the unit round three-sphere, 
a real value of $\beta$ determines a purely imaginary~$\t=\frac{\i\beta}{2\pi}$.)
The coefficient of the leading~$1/\beta^2$ term in these works is pure imaginary, and 
also does not grow as $N^2$ (it is in fact zero for non-chiral theories), therefore the 
exponential growth of states corresponding to the BH is not captured there.


One of the motivations for the current paper is to explain the exponential growth associated to the bulk black holes
from the three-dimensional point of view, which 
requires  $\arg(\tau) \neq \pi/2$ and $n_0\neq0$.\footnote{The leading order~$1/\tau^2$ behavior was found 
from similar considerations in~\cite{Choi:2018hmj}
for~$\mathcal{N}=4$ SYM with flavor chemical potentials, and in \cite{Kim:2019yrz} 
for more general gauge theories in a setting similar to ours. In this paper we follow a 
systematic, manifestly supersymmetric approach developed in~\cite{DiPietro:2014bca,DiPietro:2016ond}, 
which allows us to obtain all-order results for general gauge theories around generic rational points.}
(Note, in particular, that the $(m,n)=(1,0)$ saddle in \eqref{actionEll} given for $n_0=-1$, would 
have its leading piece a pure phase if $\mathrm{arg}(\t)=\pi/2$.) For this purpose we consider, as 
in~\cite{Cabo-Bizet:2018ehj}, a background geometry of the 
form~$S^3\times_\Omega S^1$, with $S^3$ the unit round three-sphere, 
$\rr$ the circumference of the circle, and~$\Omega$ 
a twist parameter\footnote{Similar twists had been described in slightly different contexts 
in~\cite{Kim:2009wb,Nawata:2011un}.} 
controlling the deviation of the metric from a direct product form (Equation~\eqref{4dbackgnd}).
The imaginary part of the twist parameter determines a non-zero real part 
of~$\tau$ via (Equation~\eqref{Omomrel})
\be
\t \= \frac{\i\rr}{2\pi} (1-\O) \,.
\label{eq:tauAndOmega}
\ee
As shown in~\cite{Cabo-Bizet:2018ehj}, 
the integer~$n_0$ in~\eqref{eq:n0Index} controls the 
periodicity of the fermions in this background, and~$n_0=\pm 1$ (which is naturally dual to the BH) corresponds to 
anti-periodic fermions, i.e.,~as in a Scherck-Schwarz reduction. 
In the present context we insist on supersymmetry being preserved---and that necessitates the 
turning on of other background fields under which the fermions are charged. 
In the three-dimensional background supergravity, we have a non-zero graviphoton from the fibration  
as well as non-zero auxiliary background gauge and scalar fields. As we explain in Section~\ref{sec:4dto3d}, 
the resulting configuration is effectively described by a circle of radius~$R$, which in the limit~$\rr \to 0$, $\Omega\to\infty$ with~$\t$ fixed obeys~$R \to \t$.

Now, what is the actual calculation? 
There are two types of fields in the three-dimensional functional integral---background fields 
which take constant values, and dynamical modes which fluctuate in the integral. The latter is further 
made up of light modes (with zero momentum around~$S^1$) and heavy (Kaluza-Klein) modes. 
The first step is to integrate out the heavy modes in order to obtain an effective action for the light modes. 
The integration over heavy modes also generates corrections to the coefficients of the supersymmetric
Chern-Simons terms of the non-zero background fields, see e.g.~\cite{Intriligator:2013lca,DiPietro:2014bca}.
In these calculations, we need to include, in addition to the couplings discussed 
in~\cite{DiPietro:2016ond}, the supersymmetrized RR and gravitational CS actions which were 
discussed in~\cite{Closset:2018ghr}. The effective actions of the background gauge fields turn out to 
produce precisely the singular pieces~$1/\t^2$ and~$1/\t$ in the asymptotic expansion of the index, 
as well as a constant piece. The remaining functional integral is described by an~$\CN=2$
SYM theory with a certain one-loop induced CS coupling on~$S^3$, whose partition function is 
known to agree, up to a sign, with that of pure Chern-Simons theory~\cite{Kapustin:2009kz}. 
This explains the appearance of the dynamical Chern-Simons theory in the constant term
of the asymptotic expansion.

Two technical remarks are in order. 
Firstly, recall that supersymmetry implies that the 4d superconformal index should not depend on~$\rr$
and $\Omega$ separately, but only on their combination~$\tau$ as in~\eqref{eq:tauAndOmega}. 
In~\cite{Cabo-Bizet:2018ehj} this was shown to be true in 5d gravitational variables, as well as 
through a localization computation in 4d field theory. In this paper we verify this also in 3d  effective field theory.
Secondly, the order of limits is important to have a smooth calculational set up. 
We first send~$\rr \to0$ keeping~$\Omega$ fixed, so that the three-dimensional geometry is smooth and finite. 
Then we take~$\Omega \to \infty$ at fixed~$\t$ and express the 
result in terms of~$\tau$ using~\eqref{eq:tauAndOmega}. We find there are no singularities generated in 
the latter step and thus the limiting procedure is perfectly smooth.

Finally, we repeat the same analysis as~$\t$ approaches rational points. The dimensional reduction in this
case naturally leads us to considering orbifolds of~$S^3\times S^1$, 
which, as far as we understand,
are related to the orbifolds discussed in~\cite{AhaSBTalk}. The three-dimensional calculation 
then leads to the~$1/\wt \t^2$ and~$1/\wt \t$ terms as well as a constant piece from the background 
fields, and we provide evidence that the remaining dynamical piece is the partition function 
of~$\CN=2$ SYM with a one-loop induced CS coupling on~$S^3/\mathbb{Z}_m$.

\vskip 0.4cm

\ndt {\bf Notation}. We have~$\s,\t \in \mathbb{H}$ and~$z, u \in \IC$, and 
we set~$p=\rme^{2\pi i\s}$, $q=\rme^{2 \pi \i \t}$, $\z=\rme^{2\pi\i z}$. \\
We use~$\simeq$ to mean an all-order asymptotic equality of the logarithms of the two sides.

\section{The 4d superconformal index and its asymptotic expansion \label{sec:SCI}}

We consider a four-dimensional $\CN=1$ gauge theory which flows to a superconformal fixed point. 
The theory has gauge group~$G$ (which we take to be semi-simple, and separately comment 
on the $U(N)$ case), and a number of chiral multiplets labelled by~$I$  
with R-charge $r_I$ and in the representation $\mathcal{R}_I$ of the gauge group. 
We assume $0<r_I<2$ for all chiral multiplets. 
The superconformal index for these theories on~$S^3\times S^1$
has been calculated in the Hamiltonian as well as functional integral 
formalism~\cite{Romelsberger:2005eg,Kinney:2005ej,Nawata:2011un,Assel:2014paa}, 
and the answer is expressed as an integral over the Cartan torus which we parameterize by the 
vector of gauge holonomies~$\uu = (u_1, \dots, u_{\text{rk}(G)})$, with~$u_i \in \IR/\IZ$.
The index is given by the following expression~\cite{Dolan:2008qi,Spiridonov:2009za,Spiridonov:2011hf}
\be\label{eq:pqIndex}
\mathcal I(\sigma,\tau) 
\= \int [D\uu]   \, \mathcal{Z}_{\rm vec}  (\uu;\s,\t) \, \mathcal{Z}_{\rm chi}  (\uu;\s,\t) \, .
\ee
Here we have used the measure~$D\uu = \frac{1}{|\mathcal{W}|} \prod_{i=1}^{\text{rk}(G)} \dd u_i$  
with $|\mathcal{W}|$ the order of the Weyl group of $G$. 
For~$U(N)$ we have~$D\uu = \frac{1}{N!} \prod_{i=1}^{N} \dd u_i$, while 
for~$SU(N)$ one can work with~$u_1, \dots , u_N$ subject to~$\sum_i u_i  \in \IZ$ 
and~$D\uu = \frac{1}{N!} \prod_{i=1}^{N-1} \dd u_i$.
The factors~$\mathcal{Z}_{\rm vec}$, $\mathcal{Z}_{\rm chi}$ denote the vector multiplet and
chiral multiplet contribution respectively given by
\be
\begin{split}
\mathcal{Z}_{\rm vec} (\uu;\s,\t) &\= (p;p)^{{\rm rk}(G)}(q;q)^{{\rm rk}(G)} \prod_{ \alpha_+}   
\Ge \bigl(\alpha_+\cdot u + \sigma + \tau ; \sigma , \tau \bigr)  \, 
\Ge \bigl( - \alpha_+\cdot u + \sigma + \tau ; \sigma , \tau \bigr) \, ,\\
\mathcal{Z}_{\rm chi} (\uu;\s,\t) &\= \prod_{I} \prod_{ \rho \in \mathcal{R}_I}
\Ge \Bigl(\rho\cdot u+\frac{r_I}{2} (\sigma+\tau); \sigma,\tau \Bigr) \,.
\end{split}
\label{ZvecZchi}
\ee
Here $\a_+$ runs over the set of positive roots of the gauge group $G$, 
$I$ runs over all the chiral multiplets of the theory, 
and $\rho$ is the weight of the gauge representation $\mathcal{R}_I$.  
The \emph{Pochhammer symbol} is defined by 
\be
    (\z;q) \= \prod_{k=0}^{\infty}(1- \z \, q^k) \,, \label{eq:PochDef}
\ee
and
the \emph{elliptic gamma function}~\cite{Ruijsenaars:1997,Felder}
is defined by the infinite product formula 
\be
\Ge(z;\s,\t) \= 
\prod_{j,k\ge 0}\frac{1-\,p^{j+1} \,q^{k+1} \, \z^{-1}}{1-p^{j} \,q^{k}\, \zeta } \,. 
\label{eq:GammaDef}
\ee

From now on in this section we set~$\s=\t-n_0$, and use the notation~$\Ge(z) = \Ge(z;\t,\t)$.
We have 
\be 
\CI(\t;n_0)  \=   \int [D \uu] \, \exp \bigl( -S_\text{ind}(\uu;\t)  \bigr) \,, 
\label{eq:SYMIndex} 
\ee
where the action~$S_\text{ind}(\uu) = S_\text{ind}(\uu;\t)$ is given by
\be
\begin{split}
 -S_\text{ind}(\uu)  & \=  2 \, {\rm rk}(G) \log (q;q) +  
 \sum_{ \alpha_+}  \log \bigl( \Ge \bigl(\alpha_+\cdot u +  2\tau \bigr) 
 \Ge \bigl( - \alpha_+\cdot u + 2\tau \bigr) \bigr) \\
&  \qquad  + \sum_{I} \sum_{ \rho \in \text{R}_I}
\log \Ge \bigl(\rho\cdot u+ r_I (\tau - \tfrac12 n_0) \bigr) \,.
\label{defVmicro}
\end{split}
\ee
Our goal now is to calculate the asymptotics of the function~$\CI(\t,n_0)$ as~$\t$ 
approaches a rational number or, equivalently,~$q=e^{2 \pi \i \tau}$ approaches a root of unity.
For~$\CN=4$ SYM we have 
\be 
\begin{split}
 -S_\text{ind}^{\CN=4}(\uu)  \= & 2N \log (q;q) +  3N \log \Ge \bigl(\,\tfrac13 (2\t - n_0)\bigr)   \\
& \quad + \sum_{i\neq j}  \log \Ge \bigl(u_{ij} +2\t\bigr) + 
3 \sum_{i\neq j} \log  \Ge \bigl(u_{ij} + \tfrac13 (2\t - n_0) \bigr)
\label{eq:N4Index} 
\end{split}
\ee
for~$U(N)$ and a similar expression for~$SU(N)$ with~$N$ replaced by~$N-1$.
Using the product expression~\eqref{eq:GammaDef} we see 
that for~$\CN=4$ SYM the index $\CI_N(\t;n_0)$ has the symmetry~$\t \mapsto \t+3$ for fixed~$n_0$, 
so that we can restrict our attention to, say,~$\t \in [0,3]$. Relatedly, the independent values of~$n_0$
are~$0$, $\pm 1$.  
More generally, the periodicity of~$\t$ depends on the quantization of R-charge in the theory.

Before analyzing these asymptotic limits 
we briefly discuss a slightly independent motivation to study these new limits and, relatedly, the 
origin of the number~$n_0$ in~\eqref{eq:SYMIndex}, \eqref{eq:N4Index}. 
One of the motivations in the recent developments in this subject has been to ``find the dual black hole"
in the superconformal index. In terms of the microcanonical Fourier coefficients
\be
\CI_N (\t;n_0) \= \sum_n d_N(n;n_0) \, \rme^{2 \pi \i n \t} \,,
\ee 
the problem in the context of the Cardy-like limit is to check if~$|d_N(n;n_0)|  \sim N^2 \, s(n/N^2)$  
as~$n \to \infty$~\cite{Murthy:2020rbd}.
In the canonical setting this is reflected by a corresponding asymptotic growth of the function~$\CI(\t)$ 
as~$\t$ approaches the real axis or, equivalently, as~$q = e^{2 \pi \i \t}$ approaches the unit circle. 
The leading asymptotics of the growth of microcanonical degeneracies is governed by the dominant 
singularity of~$\CI$. 
As it turns out, the index~$\CI_N(\t;0)$ of~$\CN=4$ SYM does not have any exponential growth 
as~$\t \to 0$ (the growth is power-law \cite{Ardehali:2015bla}). 
It is the asymptotic growth of~$\log \CI_N(\t;0)$ as~$\t \to \pm 1$ instead that matches the 
on-shell action of the AdS$_5$ BH (the two points giving growth of equal magnitude and opposite phases). 
From a numerical study of the microcanonical degeneracies one can deduce that this is, in fact, the 
leading growth of the index~\cite{Murthy:2020rbd}. 
In this case, noting that~$\mathcal{I}_{N}(\tau\pm1; n_0) =\mathcal{I}_{N}(\tau;n_0\mp1)$, 
we see that the leading growth can be stated as coming from the growth of the function~$\CI_N(\t,\mp1)$
as~$\t \to 0$. Actually, one finds that the growth of states at~$n_0=\pm 1$ matches the BH growth of states 
for very large classes of~$\CN=1$ SCFTs~\cite{Cabo-Bizet:2019osg,Kim:2019yrz}. 
Once we understand that the growth can come from a region with~$\Im \, \t \to 0$ but~$\Re \, \t \neq 0$,
it is perhaps more natural to set~$n_0=0$ (for which the two regions of leading growth have a symmetric
placement around~$\t=0$).
We will, nevertheless, keep~$n_0$ as an explicit parameter in the following to make contact with related literature. 

It is clear from the above discussion that one should equally well explore other points on the unit circle 
in~$q$.\footnote{The superconformal index as a function of $q$ is defined on a branched  
cover of the complex plane and one should explore the full covering space. 
For~$\CN=4$ SYM one has a three-sheeted cover and the leading 
growth occurs on two of the three sheets~\cite{Cabo-Bizet:2019osg,Kim:2019yrz,ArabiArdehali:2019orz}.}
As it turns out there is exponential growth near any root of unity consistent with~\eqref{mnasymp}, 
\eqref{actionEll}, i.e.~partial deconfinement in the sense of asymptotic growth \cite{ArabiArdehali:2019orz}. 
In the following subsections we proceed to analyze the asymptotic
behavior of the index as~$\t \to 0$ and then as~$\t$ approaches any rational number.

\subsection{Asymptotics of the index as $\tau\to0$}

In this subsection we perform an all-order asymptotic analysis of the integral~\eqref{eq:SYMIndex} 
as~$\t \to 0$. This calculation was done for~$\CN=4$ SYM recently in~\cite{GonzalezLezcano:2020yeb} 
using a saddle-point  analysis. 
Here we find the asymptotics for the general class of theories discussed in the introduction, 
using the rigorous method of~\cite{Rains:2006dfy,Ardehali:2015bla} (see in particular Section~3.1 
of \cite{Ardehali:2015bla}). 
The application in~\cite{Ardehali:2015bla} was restricted to real~$\t$ and~$n_0=0$, but the method is 
more general and we apply it to the case of complex~$\t$ and general~$n_0$.

We first calculate the all-order asymptotic expansion of the integrand. 
In order to do this we need the the asymptotic behavior of the elements in~\eqref{defVmicro}, 
namely the Pocchammer symbol and the elliptic gamma function, which 
we review in Equations~(\ref{eq:PochEst}), (\ref{eq:numEst}), \eqref{eq:denomEst}.
Using these estimates we find that in the range~$\a_+ \cdot \uu \in (-1+\delta,1-\delta)$ (for fixed small~$\delta$)
the integrand of~\eqref{eq:SYMIndex} can be written, up to exponentially suppressed corrections, as 
\be
 \exp \bigl( -S_\text{ind}(\uu;\t)  \bigr) \; \simeq \; \frac{1}{(-\i\, \tau)^{\mathrm{rk}(G)}}\, 
\prod_{\alpha_+}4\sinh^2\Bigl(\frac{\pi \alpha_+ \cdot \uu}{-\i\, \tau}\Bigr) \,  
\exp \bigl( -2\pi \i\tau E_{\mathrm{susy}} -V (\uu) \bigr)\, .
\label{eq:In0semi-simpleAsy0}
\ee
The all-order effective potential as~$\t\to 0$  is given by
\be
V(\uu) \= \frac{1}{\t^2} V_2(\uu)+ \frac{1}{\t}  V_1(\uu)+V_0(\uu) \,,
\label{defVsum}
\ee
with
\be
\begin{split}
    V_2(\uu)&\=\sum_{I, \,\rho_I}\, \frac{\i\pi}{3} \, \overline{B}_3 \bigl(\rho_I \cdot \uu  - \tfrac12 r_I n_0 \bigr) \,,\\
    V_1(\uu)&\=\sum_{I,\,\rho_I} \, \i\pi(r_I-1) \, 
    \overline{B}_2 \bigl(\rho_I \cdot \uu  - \tfrac12 r_I n_0\bigr)+
    \sum_{\alpha} \i\pi \, \Bigl((\alpha \cdot \uu)^2  + \frac{1}{6}\Bigr) \,,\\
    V_0(\uu)&\=\sum_{I,\,\rho_I} \, \i\pi \Bigl( (r_I-1)^2-\frac{1}{6} \Bigr) \, 
    \overline{B}_1  \bigl( \rho_I \cdot \uu  - \tfrac12 r_I n_0\bigr) \,,
\label{eq:defVs}
\end{split}
\ee
where~$\a$ runs over all the roots of $G$ including the rk$(G)$ zero roots,~$I$ runs over all the 
chiral multiplets, and~$\rho_I$ runs over all the weights of the representation~$\mathcal{R}_I$.
Note that in~\eqref{eq:In0semi-simpleAsy0} we have separated
the supersymmetric Casimir energy given by~\cite{Assel:2015nca}
\be
E_{\mathrm{susy}} \= \frac{1}{6} \, \mathrm{Tr}R^3 -\frac{1}{12} \, \mathrm{Tr}R\,.
\label{Esusy}
\ee

We make a brief comparison to~\cite{Cabo-Bizet:2019osg} in which the singular pieces were studied. 
The potential~$V_2$ in~\eqref{eq:defVs} coincides (up a multiplicative~$-\i\pi/6$ factor) with 
the~$V_2$ studied in~\cite{Cabo-Bizet:2019osg}. 
At finite $\uu$, the sinh$^2(\frac{\pi\alpha_+\cdot\uu}{-\i\tau})$ factors in~\eqref{eq:In0semi-simpleAsy0} 
also contribute to~$\rO(1/\t)$. Including this piece in~$V_1$ renormalizes it to~$V^r_1$ 
as\footnote{Instead of using~\eqref{eq:denomEst} for vector multiplet gammas and then 
simplifying the sinh term for finite~$\uu$ as above, we could alternatively use~\eqref{eq:numEst} 
for vector multiplet gammas (assuming~$\alpha_+\cdot\uu\notin\mathbb{Z}$) and get~$V_1^r$ 
directly as in~\cite{Cabo-Bizet:2019osg}.}
\begin{equation}
    V_1(\uu)\to V^r_1(\uu) \=\sum_{I,\,\rho_I} \, \i\pi(r_I-1) \, 
    \overline{B}_2 \bigl(\rho_I \cdot \uu  - \tfrac12 r_I n_0\bigr)+
    \sum_{\alpha} \i\pi \, \overline{B}_2 \bigl(\alpha \cdot \uu \bigr) \,.\label{eq:V1ren}
\end{equation}
The potential~$V^r_1$ coincides (up to a multiplicative $-\i\pi$ factor) with the~$V_1$ studied in~\cite{Cabo-Bizet:2019osg}. 
In our treatment below we keep the sinh$^2$ factors separate and place them in the ``dynamical measure" 
\be
\frac{D\uu}{(-\i\tau)^{\mathrm{rk}(G)}}\, 
\prod_{\alpha_+}4\sinh^2\Bigl(\frac{\pi \alpha_+ \cdot \uu}{-\i\tau}\Bigr) \,. 
\ee
Compared to~\cite{Cabo-Bizet:2019osg}, here we also include the $\rO(\tau^0)$ piece 
corresponding to $\exp(-V_0)$. 
Finally, the $\rO(\tau)$ piece of the exponent is determined by the 
supersymmetric Casimir energy and, notably, there are no $\rO(\tau^2)$ or higher 
corrections in the perturbative effective potential.

We now investigate the local behavior of the potential near~$\uu =0$. 
The potentials~$V_{2}, V_1, V_0$ are piecewise polynomials,
and using~$\overline{B}'_j=j \overline{B}_{j-1}$ we obtain their Taylor expansion near~$\uu=0$ as 
\be
\begin{split}
    V_2(\uu) &\=\sum_{I,\,\rho_I} \Bigl(\frac{\i\pi}{3} \, \overline{B}_3 \bigl(- \tfrac12 r_I \, n_0 \bigr) \+
    \i\pi \, \overline{B}_2 \bigl(-\tfrac12 r_I \, n_0 \bigr)\rho_I\cdot\uu
    \+2\pi \i \, \overline{B}_1\bigl(-\tfrac12 r_I \, n_0 \bigr)\frac{(\rho_I\cdot\uu)^2}{2} \Bigr)\\
    &\qquad \+\sum_{I,\,\rho_I} \, 2\pi\i \, \frac{(\rho_I\cdot\uu)^3}{3!}\, ,\\
    V_1(\uu) &\=\sum_{I,\,\rho_I} \, \i\pi(r_I-1) \, 
    \overline{B}_2\bigl(-\tfrac12 r_I \, n_0\bigr)\+\frac{\i\pi}{6}\text{dim}G\+
    \sum_{\rho_I}2\pi\i \, (r_I-1) \, \overline{B}_1(-\tfrac12 r_I \, n_0)\, \rho_I\cdot\uu\\ 
    &\qquad \+\sum_{I,\,\rho_I} \, 2\pi\i \,(r_I-1) \, \frac{(\rho_I\cdot\uu)^2}{2}
    \+\sum_{\alpha} \i\pi \, (\alpha \cdot \uu)^2,\\
    V_0(\uu) &\=\sum_{I,\,\rho_I} \, \i\pi \Bigl((r_I-1)^2-\frac{1}{6} \Bigr)
    \overline{B}_1\bigl(-\tfrac12 r_I \, n_0\bigr)\\
    &\qquad \+\sum_{I,\,\rho_I} \, \i\pi \Bigl((r_I-1)^2-\frac{1}{6} \Bigl)\rho_I\cdot\uu.
\end{split}
\ee
Importantly, the second lines of $V_{2}$, $V_1$, $V_0$ above vanish due to gauge$^3$, 
$U(1)_R$-gauge$^2$, and $U(1)_R^2$-gauge and gravity$^2$-gauge anomaly cancellations. 
Therefore $V_2$ is actually piecewise quadratic in $\uu$, while $V_1$ is piecewise linear 
and $V_0$ is piecewise constant. This is similar to Section~3.1 of~\cite{Ardehali:2015bla}.

The leading asymptotic behavior of~$V$~as $\tau\to0$ is determined by~$V_2$.  
In order to obtain a local minimum of~$\text{Re}(V)$ at $\uu=0$, we want 
(i) the linear term in $V_2$ to vanish, and (ii) the quadratic term to be on the negative (respectively positive) 
imaginary axis for~$\mathrm{arg}(\tau)-\frac{\pi}{2}>0$ (respectively~$\mathrm{arg}(\tau)-\frac{\pi}{2}<0$). 
As found in~\cite{Cabo-Bizet:2019osg}, we can achieve both 
of these requirements in any theory in which~$0<r_I<2$ by specializing to~$n_0=\pm1$. 
Explicitly, for $n_0=\pm1$ we can use the fact that for~$|x|<1$ we 
have~$\overline{B}_2(x)=x^2-|x|+\frac{1}{6}$ to 
deduce that the linear term in~$V_2$ is equal to
\be
\sum_{I,\,\rho_I}\, \frac{\i\pi}{4} \, \Bigl( (r_I-1)^2-\frac{1}{3} \Bigl) \, \rho_I\cdot\uu \,,
\ee
which vanishes thanks to the~$U(1)_R^2$-gauge and gravity$^2$-gauge anomaly cancellations. 
Similarly we can use the fact that for~$0<|x|<1$ we have~$\overline{B}_1(x)=x-\frac{\mathrm{sign}(x)}{2}$ 
to deduce that the quadratic term in $V_2$ is equal to
\be
-\i\pi n_0 \, \sum_{I,\,\rho_I} \, (r_I-1)\frac{(\rho_I\cdot\uu)^2}{2}\=\i\pi n_0 \, 
\sum_{\alpha} \, \frac{(\alpha\cdot\uu)^2}{2} \,,
\ee
where the equality follows from~$U(1)_R$-gauge$^2$ anomaly cancellation, and we have 
used that~$\mathrm{sign}(n_0)=n_0$ for~$n_0=\pm1$.
This quadratic piece is on the positive (respectively negative) imaginary axis for $n_0=+1$ 
(respectively $n_0=-1$). In this manner we see that~$\uu=0$ is a local minimum of~$\text{Re}(V)$. 
Therefore in the rest of this subsection 
{we focus on~$n_0=\pm1$, and take~$\mathrm{arg}(\tau)-\frac{\pi}{2}$ to have the 
opposite sign to~$n_0$}, i.e.~$n_0\, (\mathrm{arg}(\tau)-\frac{\pi}{2})<0$.

Using the explicit expressions of~$\overline{B}_{1,2,3}(x)$ in the range~$0<|x|<1$, and using the 
anomaly cancellation conditions, the potentials~$V_2$,~$V_1$,~$V_0$ simplify, for~$n_0=\pm1$, to
\be
\begin{split}
    V_2(\uu)&\=-\frac{\i\pi n_0}{24}\, \bigl(\mathrm{Tr}R^3-\mathrm{Tr}R\bigr) + \i\pi n_0 \, 
    \sum_{\alpha}\frac{(\alpha\cdot\uu)^2}{2} \,,\\
    V_1(\uu)&\=\frac{\i\pi}{12} \, \bigl(3\mathrm{Tr}R^3-\mathrm{Tr}R\bigr) \,,\\
    V_0(\uu)&\=\sum_{I,\,\rho_I}\, \i\pi \, \Bigl(\frac{r_I-1}{6} - (r_I-1)^3\Bigr)
    \frac{n_0}{2} \,.
\label{eq:simplifiedVs}
\end{split}
\ee
Note that, as a bonus, $V_1$ also becomes independent of $\uu$ for $n_0=\pm1$ and 
small enough~$\uu$.\footnote{We will shortly interpret the quadratic term in $V_2$ as inducing a 
Chern-Simons type coupling in the integrand. If the linear term in $V_1$ were present, 
it would similarly induce an FI parameter in the integrand. While this is impossible for semi-simple 
gauge theories near~$\uu=0$ which we are focussing on in this section, there are cases of 
semi-simple gauge theories in which one must expand 
around~$\uu\neq0$ and as a result one finds the measure of an abelian gauge theory in the integrand, 
where such induced FI parameters do arise. See Section~3.3.1 of~\cite{Ardehali:2015bla} where 
the ISS model displaying an SU(2)$\to U(1)$ breaking pattern with an induced FI parameter in 
the~$\tau\to0$ limit is discussed, and see Appendix~A of~\cite{ArabiArdehali:2019zac} where that 
induced FI parameter is given an effective field theory explanation.}

We now consider a small neighborhood $\mathfrak{h}_{cl}^\epsilon$ around $\uu=0$, 
defined by the cutoff $|u_j|<\epsilon$, whose contribution to the index is 
\be
\mathcal{I}(\tau;n_0)_{|u_j|<\epsilon} \; \simeq \;
\rme^{-2\pi \i\tau E_{\mathrm{susy}}}\int_{\mathfrak{h}_{cl}^\epsilon} \frac{D\uu}{(-\i\tau)^{\mathrm{rk}(G)}}\, 
\prod_{\alpha_+}4\sinh^2\Bigl(\frac{\pi \alpha_+ \cdot \uu}{-\i\tau}\Bigr) \,  
\exp \bigl( -V(\uu) \bigr)\, .
\label{eq:In0semi-simpleAsy1}
\ee
From the above discussion we have that 
\be
\mathcal{I}(\tau;n_0=\pm1)_{|u_j|<\epsilon} \; \simeq \; \rme^{-2\pi \i\tau E_{\mathrm{susy}}} \, 
Z^{\text{bgnd}}(\tau;n_0) \  Z^{\text{dyn}}_\epsilon(\tau;n_0)\,,
\label{eq:In0semi-simpleAsy2}
\ee
where the \emph{background piece} is 
\be
\begin{split}
& Z^{\text{bgnd}}(\tau;n_0)  \= \\ 
& \quad \exp \Bigl( \frac{\i\pi\, n_0}{24\tau^2}(\mathrm{Tr}R^3-\mathrm{Tr}R)
-\frac{\i\pi}{12\tau}(3\mathrm{Tr}R^3-\mathrm{Tr}R) + 
\sum_{\rho_I}\frac{\i\pi\, n_0}{2} \bigl((r_I-1)^3-\tfrac16(r_I-1) \bigr) \Bigr)
\end{split}
\ee
and the \emph{dynamical piece} is
\be
Z^{\text{dyn}}_\epsilon(\tau;n_0)=\int_{\mathfrak{h}_{cl}^\epsilon} \frac{D\uu}{(-\i\, \tau)^{\mathrm{rk}(G)}}\,
\prod_{\alpha_+}4\sinh^2\Bigl(\frac{\pi \alpha_+ \cdot \uu}{-\i\,\tau}\Bigr) \,  
\exp \Bigl(\frac{\i\pi n_0}{2}\sum_\alpha \bigl(\frac{\alpha\cdot\uu}{-\i\,\tau} \bigr)^2 \Bigr) \,.\label{eq:ZdynComplex}
\ee
Here we suppress the dependence of these functions on the gauge group and the matter content.

To simplify $Z_\epsilon^{\text{dyn}}$ further, we first define $x_j=\frac{u_j}{-\i\tau}$, 
so that the integral becomes along straight contours from 
$x_j=-\frac{\epsilon}{-\i\tau}$ to $x_j=+\frac{\epsilon}{-\i\tau}$. 
With our choice of~$n_0$ and~arg$(\t)$, the integrand is 
locally exponentially suppressed away from $\uu=0$, so we can complete the tails of the contours 
along straight lines to infinity (i.e.~send~$\epsilon\to+\infty$) introducing only exponentially small error. 
The contours make an angle~$\frac{\pi}{2}-\mathrm{arg}(\tau)$ with the positive real axis. 
However, observing that (i)~the integrand is exponentially suppressed as $|x_j|\to\infty$, and (ii) 
there are no poles between the contour of $x_j$ and the real axis, we can deform the contours 
back to the real axis. We thus obtain, with~$ \ux = (x_1,\dots,x_n)$ 
\be
Z^{\text{dyn}}_\epsilon(\tau;n_0) \;\simeq \; \int_{-\infty}^\infty D \ux \,
\prod_{\alpha_+}4\sinh^2(\pi \alpha_+ \cdot \ux) \,  
\exp \Bigl( \frac{\i\pi n_0}{2}\sum_\alpha (\alpha\cdot \ux )^2 \Bigr) \; =: \; Z^{\text{dyn}}(n_0) \,.\label{eq:ZdynS3}
\ee
As noted in~\cite{GonzalezLezcano:2020yeb} for~$\CN=4$ SYM,
and in \cite{Amariti:2020jyx,Amariti:2021ubd} for more general groups, 
$Z^{\text{dyn}}$ is related to the
partition function of pure Chern-Simons theory~\cite{Witten:1988hf} on~$S^3$ as
\be
Z^{\text{dyn}} ( n_0)\=(-1)^{(\text{dim}G-\text{rk}G)/2} \, Z^{\text{CS}}(k_{ij}) \,,
\label{dynCSrel}
\ee
with the gauge group implicit and the same on both sides, and with Chern-Simons coupling given by 
\be
k_{ij}\=-\frac{n_0}{2} \,\sum_\alpha \alpha_i \, \alpha_j \,.
\ee
Notice that~$Z^{\text{dyn}} ( n_0)$ is independent of~$\t$. The tails completion 
(i.e.~sending~$\epsilon\to\infty$) and contour deformation mentioned above 
removed the~$\tau$-dependence of~$Z_\epsilon^{\mathrm{dyn}}$ at the cost of an exponentially small error.

The considerations of three-dimensional effective field theory in the next section show 
that~$Z^{\text{dyn}}$ arises naturally in fact as the partition function of three-dimensional~$\CN=2$ 
gauge theory on~$S^3$ with the same gauge group and the same CS coupling. (The latter is 
well-known to be related  to~$Z^{\text{CS}}$ exactly as in~\eqref{dynCSrel}.)

\vskip 0.4cm

The above analysis was local around~$\uu=0$. 
We now focus on SU($N$) $\CN=4$ SYM for which we know that~$\uu=0$ is a global minimum 
of the leading potential~$V_2/\t^2$ for $n_0=\pm1$ and 
$n_0(\mathrm{arg}(\t)-\frac{\pi}{2})<0$.\footnote{This was shown 
in~\cite{Cabo-Bizet:2019osg,Kim:2019yrz} where experimental evidence that this is true for a large class 
of theories was also discussed.} 
However, this is not the only global minimum---there are~$N$ isolated global mimima 
labelled by~$k = 1, \dots ,N$ 
which are related to~$u_j=0$ by center symmetry, namely the points~$u_j=(k-1)/N$, 
$k=1, \dots, N$ \cite{ArabiArdehali:2019tdm}. 
Upon summing over these minima we obtain
\be 
 \CI_N(\t;n_0) \; \simeq \;  N \rme^{-2\pi\i\tau E_{\text{susy}}}  \, Z^\text{bgnd}(\t;n_0) \,  Z^\text{dyn}(n_0) \,. 
\label{eq:almostThere}
\ee
The factor of $N$ arises from the sum over~$N$ minima as explained above. 
The other three factors can be calculated by 
specializing our general discussion to this case: 
\bea
E_{\mathrm{susy}}& \= & \frac{4}{27}\big(N^2-1\big) \,, \\
Z^\text{bgnd}(\t;n_0) & \= &   \exp\left(-\frac{\i\pi}{\tau^2}(N^2-1)
    \biggl(\Bigl( \frac{-n_0+2\tau}{3} \Bigr)^3 + 
    \frac{5n_0\tau^2}{12} \biggr)+2\pi\i\tau\cdot\frac{4}{27}\big(N^2-1\big)\right) \,, 
    \qquad\qquad 
\label{eq:Zback} \\
Z^\text{dyn}(n_0)  & \=  &
 \int_{-\infty}^\infty D \ux \,    \prod_{i<j} 4 \sinh^2 \bigl(\pi x_{ij} \bigr) \,
    \exp \Bigl(  \i\pi n_0 N \sum_{j=1}^N x_j^2 \, \Bigr)  \,.
    \label{Zdyn}
\eea
In this case the matrix of Chern-Simons couplings reduces to a single level ($k_{ij} = k \delta_{ij}$), 
and we have
\be
Z^\text{dyn}(n_0) \= (-1)^{N(N-1)/2} \, Z^{\text{CS}}(k) \,, \qquad k \= -n_0 N \,.
\label{ZdynCS}
\ee

For $n_0=\pm1$ the SU($N$) Chern-Simons partition was found in \cite{GonzalezLezcano:2020yeb} to simplify to
\begin{equation}
Z^\text{CS}(-n_0 N) \=(-1)^{N(N-1)/2}\exp \bigl(5 \i\pi \,n_0 \ (N^2-1)/12 \bigr) \,.    
\end{equation}
Upon combining this equation with~\eqref{ZdynCS} and \eqref{eq:almostThere}, 
we obtain\footnote{For comparison with~\cite{Lezcano:2019pae}, we set~$\xi_a^{\text{there}}=-n_0^{\text{here}}/3$.
The result in that paper contains the number $\eta\in\{-1,+1\}$. This is related to our~$n_0$ as 
$\eta=6\overline{B}_1(-\frac{n_0}{3})$. For~$n_0 = \pm 1$, a simple calculation shows that $\eta=n_0$.}
\begin{equation}
\begin{split}
   \CI_N(\t;n_0) \; \simeq \; 
    N \, \exp \biggl(-\frac{\i\pi}{\tau^2} \, (N^2-1) \Bigl(\frac{-n_0+2\tau}{3} \Bigr)^3 \biggr) \,.
    \label{eq:N=4indexAsyFinal}
    \end{split}
\end{equation}
The analogous result for the case with $U$($N$) gauge group is obtained by adding the 
contribution of a decoupled~$U$(1) $\mathcal{N}=4$ multiplet to that of the SU($N$) theory:
\begin{equation}
\begin{split}
    \mathcal{I}^{U(N)}(\t;n_0) \; \simeq \;  
    N \, \frac{1}{-\i\tau} \, \exp \biggl( -\frac{\i\pi}{\tau^2} \, N^2 \Bigl(\frac{-n_0+2\tau}{3} \Bigr)^3
    -\frac{5\pi \i \, n_0}{12} \biggr) \,.
    \label{eq:N=4indexAsyFinalU(N)}
    \end{split}
\end{equation}

This finishes the discussion of our methods illustrated in the special case~$\tau\to0$. 
Before moving on to the more general case of rational points, we make a few technical remarks.

Firstly, since we are analyzing the index by estimating its integrand, we need uniform estimates. 
For~$n_0=\pm1$, the estimate~\eqref{eq:numEst} when applied to the chiral multiplet gamma functions 
gives uniformly valid asymptotics near~$\uu=0$, because the~$-n_0 \, r_I/2$ shift in the argument takes 
us safely into the domain of validity 
of~\eqref{eq:numEst}. For the vector multiplet gamma functions, however, there is no finite shift in the 
argument, so~\eqref{eq:numEst} does not apply uniformly around~$\uu=0$. We had to use 
instead~\eqref{eq:denomEst} to obtain uniform asymptotics near~$\uu=0$ for the vector multiplet gammas.

Secondly, we emphasize that our asymptotic analysis is essentially real-analytic 
(as in~\cite{Rains:2006dfy,Ardehali:2015bla}). We only appeal to complex-analytic tools 
(specifically, contour deformation), after having done the asymptotic analysis, 
to simplify the final answer for~$Z_\epsilon^{\mathrm{dyn}}$ 
in~\eqref{eq:ZdynComplex} to the more familiar form~\eqref{eq:ZdynS3}.

Thirdly, we note that when actually doing the saddle-point analysis, one finds that the dominant 
holonomy configurations spread into the complex plane, as in the analysis 
of~\cite{Cabo-Bizet:2019eaf,Benini:2018ywd,GonzalezLezcano:2020yeb}. 
Upon taking the~$\tau\to0$ limit the spreading shrinks, and the answers from those approaches 
indeed agree with our results.

\subsection{Asymptotics of the index as $\tau\to\mathbb{Q}$}

We now study the index~\eqref{eq:SYMIndex} in the limit 
\begin{equation}
     \wt \tau \; \equiv \; m\tau+n \to 0 \,, 
\label{eq:rationalLim}
\end{equation}
with $m,n$ relatively prime, keeping~$\mathrm{arg}(\wt \tau)$ away from integer multiples 
of~$\pi/2$. 

As in the previous subsection we first calculate the all-order asymptotics of the integrand. 
The required small-$\wt\t$ estimates for the Pocchammer symbol and the elliptic gamma 
function are given in Equations~(\ref{eq:PochRationalEst}), (\ref{eq:gammaRationalEstWithR}), 
\eqref{eq:denomEstRational}.
Using these estimates we find that in the 
range~$\a_+ \cdot \uu \in (-\frac{1}{m}+\delta,\frac{1}{m}-\delta)$, for some fixed small~$\delta$,
the integrand of~\eqref{eq:SYMIndex} can be written up to exponentially suppressed corrections as 
\be
 \exp \bigl( -S_\text{ind}(\uu;\t)  \bigr) \; \simeq \; \frac{1}{(-\i\, \wt\tau)^{\mathrm{rk}(G)}}\, 
\prod_{\alpha_+}4\sinh^2\Bigl(\frac{\pi \alpha_+ \cdot \uu}{-\i\, \wt\tau}\Bigr) \,  
\exp \bigl( -2\pi \i \, \wt\tau \, \frac{E_{\mathrm{susy}}}{m} -\frac{\wt V (\uu)}{m} \bigr)\, .
\label{eq:In0semi-simpleAsyRat0}
\ee
The all-order effective potential as~$\wt\t\to 0$  is given by
\be
\wt V(\uu) \= \frac{1}{\wt\t^2} \wt V_2(\uu)+ \frac{1}{\wt\t}  \wt V_1(\uu)+\wt V_0(\uu) \,,
\label{defVsumRat}
\ee
with
\be
\begin{split}
    \wt V_2(\uu)&\=\sum_{I, \,\rho_I}\, \frac{\i\pi}{3} \, \overline{B}_3 \bigl(m\rho_I \cdot \uu +m\xi_I \bigr) \,,\\
    \wt V_1(\uu)&\=\sum_{I,\,\rho_I} \, \i\pi(r_I-1) \, 
    \overline{B}_2 \bigl(m\rho_I \cdot \uu  +m\xi_I\bigr)+
    \sum_{\alpha} \i\pi \, \Bigl((m\alpha \cdot \uu)^2  + \frac{1}{6}\Bigr) \,,\\
    \wt V_0(\uu)&\=-2\pi\i\cdot\mathrm{dim}(G)\, s(n,m)+\sum_{I,\,\rho_I} \,2\pi\i\, C(m,n,\rho_I\cdot\uu-\frac{n_0}{2}r_I,r_I) \,,
\label{eq:defVsRat}
\end{split}
\ee
where~$\a$ runs over all the roots of $G$ including the rk$(G)$ zero roots,~$I$ runs over all the 
chiral multiplets, and~$\rho_I$ runs over all the weights of the representation~$\mathcal{R}_I$. 
Here~$s(n,m)$ is the Dedekind sum defined in~\eqref{eq:ourDedekind Sum} and the function~$C(m,n,r,z)$ is defined in~\eqref{eq:C(m,n,r,z)}. 
We have defined
\begin{equation}
    \xi_I\defeq-\frac{r_I}{2}\big(n_0+\frac{2n}{m}\big),\label{eq:defXiGen}
\end{equation}
to emphasize an analogy with the analysis in~\cite{ArabiArdehali:2019orz,ArabiArdehali:2019tdm} 
of the index with flavor chemical potential~$\xi$, although we do not have flavor fugacities 
in our problem. Note also that we have separated the supersymmetric Casimir energy 
in~\eqref{eq:In0semi-simpleAsyRat0} as in the~$\t \to 0$ case.

Next, as in the previous subsection we expand the potentials near~$\uu=0$. 
Anomaly cancellations again lead to simplifications, 
but here we further assume the theory is non-chiral (i.e. that~$\rho_I$ come 
in pairs of opposite sign) so that the answer takes a particularly simple form. 
Analogously to~\eqref{eq:simplifiedVs} we obtain
\be
\begin{split}
    \wt V_2(\uu) &\=\sum_{I,\,\rho_I} \Bigl(\frac{\i\pi}{3} \, \overline{B}_3 \bigl(m\xi_I \bigr) 
    \+2\pi \i \, \overline{B}_1\bigl(m\xi_I \bigr)\frac{(m\rho_I\cdot\uu)^2}{2} \Bigr) \,,\\
    \wt V_1(\uu) &\=\sum_{I,\,\rho_I} \, \i\pi(r_I-1) \, 
    \overline{B}_2\bigl(m\xi_I\bigr)\+\frac{\i\pi}{6}\text{dim}(G)\,,\\
    \wt V_0(\uu) &\=\wt V_0(0)\= -2\pi\i \, \mathrm{dim}(G)\, s(n,m)
    +\sum_{I,\,\rho_I} \,2\pi\i\, C(m,n,-\frac{n_0}{2}r_I,r_I) \,.
\label{eq:simplifiedVtildes}
\end{split}
\ee

Next we focus on a small neighborhood~$\mathfrak{h}_{cl}^\epsilon$ around~$\uu=0$, 
defined by the cutoff~$|u_j|<\epsilon$, whose contribution to the index as~$\wt\t\to0$ is 
\be
\mathcal{I}(\tau;n_0)_{|u_j|<\epsilon} \; \simeq \;
\rme^{-2\pi \i\wt\tau \frac{E_{\mathrm{susy}}}{m}}
\int_{\mathfrak{h}_{cl}^\epsilon} \frac{D\uu}{(-\i\wt\tau)^{\mathrm{rk}(G)}}\, 
\prod_{\alpha_+}4\sinh^2\Bigl(\frac{\pi \alpha_+ \cdot \uu}{-\i\wt\tau}\Bigr) \,  
\exp \bigl( -\frac{\wt V(\uu)}{m} \bigr)\, .
\label{eq:In0semi-simpleAsyRat1}
\ee
Upon putting the above discussion together we obtain
\be
\mathcal{I}(\tau;n_0)_{|u_j|<\epsilon} \; \simeq \; \rme^{-2\pi \i\wt\tau \frac{E_{\mathrm{susy}}}{m}} \, 
Z^{\text{bgnd}}(\tau;m,n,n_0) \  Z_\epsilon^{\text{dyn}}(\tau;m,n,n_0)\,,
\label{eq:In0semi-simpleAsyRat2}
\ee
where the \emph{background piece} is 
\be
\begin{split}
& Z^{\text{bgnd}}(\tau;m,n,n_0)  \= \\ 
& \quad \exp \Bigl( -\frac{1}{m\wt\t^2}\sum_{I,\,\rho_I} \frac{\i\pi}{3} \, \overline{B}_3 \bigl(m\xi_I \bigr)
-\frac{1}{m\wt\t}\sum_{I,\,\rho_I} \, \i\pi(r_I-1) \, 
    \overline{B}_2\bigl(m\xi_I\bigr)\+\frac{\i\pi}{6}\text{dim}(G) + 
\wt V_0(0) \Bigr)\,,
\end{split}
\ee
and the \emph{dynamical piece} is
\be
Z_\epsilon^{\text{dyn}}(\tau;m,n,n_0)=\int_{\mathfrak{h}_{cl}^\epsilon} \frac{D\uu}{(-\i\, \wt\tau)^{\mathrm{rk}(G)}}\,
\prod_{\alpha_+}4\sinh^2\Bigl(\frac{\pi \alpha_+ \cdot \uu}{-\i\,\wt\tau}\Bigr) \,  
\exp \Bigl(+\frac{\i\pi}{m} \sum_{I,\,\rho_I}  \overline{B}_1 \bigl(m\xi_I \bigr)\bigl(\frac{m\rho_I\cdot\uu}{-\i\,\tau} \bigr)^2 \Bigr) \,.
\ee
Upon defining the rescaled variable~$x_j=\frac{u_j}{-\i\wt\tau}$, 
we recognize~$Z_\epsilon^{\text{dyn}}(\tau;m,n,n_0)$ as the CS partition function 
on~$S^3$ with gauge group~$G$  and 
level~$k^{ij}\=-\frac{1}{m}\sum_{I,\rho_I}\overline{B}_1 \bigl(m\xi_I \bigr) \rho_I^i \rho_I^j$. 
We will see momentarily that it is more natural to define the rescaled variable 
as~$x_j=\frac{mu_j}{-\i\wt\tau}$. Upon tails completion and deforming the integration 
contour we obtain
\be
\begin{split}
Z_\epsilon^{\text{dyn}}(\tau;m,n,n_0)\; &\simeq \; m^{-\mathrm{rk}(G)}\int_{-\infty}^{\infty} D\ux\,
\prod_{\alpha_+}4\sinh^2\Bigl(\frac{\pi \alpha_+ \cdot \ux}{m}\Bigr) \,  
\exp \Bigl(+\frac{\i\pi}{m} \sum_{I,\,\rho_I}  
\overline{B}_1 \bigl(m\xi_I \bigr)\bigl(\rho_I\cdot\ux \bigr)^2 \Bigr)\\
&=:\; m^{-\mathrm{rk}(G)}\, Z_{\underline{0}}^{\text{dyn}}(m,n,n_0)\,.\label{eq:Z0Rational}
\end{split}
\ee
Up to the overall $m^{-\mathrm{rk}(G)}$ factor, this coincides~\cite{Gang:2019juz} with the topologically 
trivial sector of the~$S^3/\mathbb{Z}_m$ partition function of $\mathcal{N}=2$ SYM with Chern-Simons coupling
\be
k^{ij}\=-\sum_{I,\rho_I}\overline{B}_1 \bigl(m\xi_I \bigr) \,  \rho_I^i \, \rho_I^j \,.
\label{eq:CScouplingRat}
\ee

While the explicit expression for the dominant potential~$\wt V_2$ in~\eqref{eq:defVsRat} 
was derived in a neighborhood $(-\frac{1}{m}+\delta,\frac{1}{m}-\delta)$ of~$\uu=0$, it is 
actually correct more generally, because it follows from~\eqref{eq:gammaRationalEstWithR} which 
we can use as long as $m\rho_I \cdot \uu +m\xi_I\notin\mathbb{Z}$. Moreover, since~$\rho^j_I$ 
are integers and~$u_j$ appears in~$\wt V_2$ in the combination~$m\,\rho_I\cdot\uu$, the 
1-periodicity of~$\wt V_2$  implies that any holonomy configuration with~$u_{j}$ a multiple 
of~$1/m$ gives the same leading asymptotics as the~$\uu=0$ configuration. 
In the SU($N$) case these non-trivial holonomy configurations correspond to
\be
\uu=\frac{1}{m} \, \um \= \bigl(\frac{m_1}{m} \,,\dots,\frac{m_N}{m} \bigr) \, ,
\label{eq:nontrivialSectors}
\ee
with~$m_j\in\mathbb{Z}/m\mathbb{Z}$, and~$\sum_{j=1}^N m_j=0$ (mod $m$).

For~$n=1$, we can use the estimate~\eqref{eq:denomEstRational0.75} for the vector multiplet 
gamma functions to compute the contribution of the saddles~\eqref{eq:nontrivialSectors}. 
The result is similar to~\eqref{eq:In0semi-simpleAsyRat1}, with the same~$\wt V_{2,1}$ and 
the same SUSY Casimir piece, but with the dynamical piece modified to (modulo an overall constant factor)
\be
\begin{split}
Z_{\epsilon_{\underline{m}}}^{\text{dyn}}(\tau;m,n,n_0)\; &\simeq \; \int_{-\infty}^{\infty} D\ux'\,
\prod_{\alpha_+}4\sinh^2\Bigl(\frac{\pi \alpha_+ \cdot (\ux'+\i\um)}{m}\Bigr) \,  
\exp \Bigl(+\frac{\i\pi}{m} \sum_{I,\,\rho_I}  
\overline{B}_1 \bigl(m\xi_I \bigr)\bigl(\rho_I\cdot\ux' \bigr)^2 \Bigr)\\
&=:\; Z_{\underline{m}}^{\text{dyn}}(m,n,n_0)\,,\label{eq:Z0RationalNonTrivialSect}
\end{split}
\ee
where~$\epsilon_{\um}$ indicates that we are considering the contribution from a 
neighborhood~$|u_j-m_j/m|<\epsilon$, and the re-scaled integration variable arises 
as~$x_j'=\frac{m(u_j-m_j/m)}{-\i\wt\t}$. This coincides (again up to an overall constant factor) 
with the topologically non-trivial sector of the partition function of~SU($N$) Chern-Simons theory 
with coupling~\eqref{eq:CScouplingRat} on the lens space~$L(m,-1)$~\cite{Gang:2019juz}.

We expect that similarly for general~$n$, including the contribution of the non-trivial 
saddles~\eqref{eq:nontrivialSectors} to the index would complete~$Z_{\underline{0}}^{\text{dyn}}(m,n,n_0)$ 
to the full~$S^3/\mathbb{Z}_m$ partition function, including all the topologically 
non-trivial sectors. We motivate this expectation 
further from an EFT perspective in the next section where we also present the ($n$-dependent) 
action of~$\mathbb{Z}_m$ on the~$S^3$. The explicit demonstration, which we leave to future work, 
requires generalizing the estimate~\eqref{eq:denomEstRational0.75} to arbitrary~$n$, and improving 
it to include the overall constant.

The above analysis was local in nature: we considered the contribution to the index from only a 
small neighborhood of~$\uu=0$. We now study the specific case of SU($N$)~$\mathcal{N}=4$ 
SYM for which we present a global picture of the dominant holonomy configurations. 
Note that in the previous subsection rather than performing the global analysis from scratch 
we had borrowed the result of~\cite{Kim:2019yrz,Cabo-Bizet:2019osg} that in a certain 
domain of parameters ($n_0(\mathrm{arg}(\t)-\pi/2)<0$) the~$\uu=0$ configuration is 
globally dominant (see Section~\ref{sec:Ccenter} for the complementary domain).

\subsubsection*{The global structure of the leading potential for $\mathcal{N}=4$ SYM}

For SU($N$) $\mathcal{N}=4$ theory the potential~$\wt V_2$ reads
\begin{equation}
    \begin{split}
    \wt V_2 (\uu;\xi) \= 
    \frac{\i\pi}{3} \times 3 \; \Bigl((N-1)\overline{B}_3(m\xi)+
    \sum_{i<j} \bigl( \overline{B}_3(m\xi+m u_{ij})+ \overline{B}_3 (m\xi + m u_{ji}) \bigr)\Bigr) \,,
    \label{eq:QhDef}
    \end{split}
\end{equation}
where the factor of 3 comes from the sum over three chiral multiplets, and 
with
\be
\xi \= - \tfrac13 \bigl(n_0 + \tfrac{2n}{m} \bigr) \,.
\label{defxi}
\ee 
As mentioned below \eqref{eq:CScouplingRat} the expression \eqref{eq:QhDef} 
applies as long as $u_{ij}+\xi_I$ avoid $\frac{\mathbb{Z}}{m}$.

\begin{figure}[h]
\centering
    \includegraphics[scale=.6]{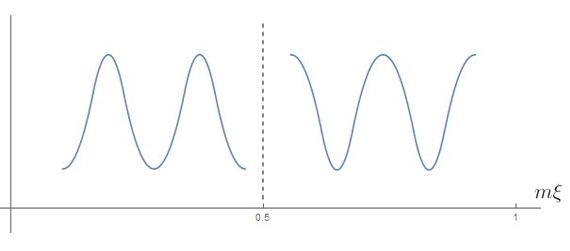}
\caption{The catastrophic behavior of $V^Q(u_{ij})$, drawn over the 
range~$mu_{ij}\in(-1,1)$, for $\mathrm{arg}\widetilde{\tau}>\frac{\pi}{2}$. 
The control parameter $m\xi$ determines the M or W type behavior. \label{fig:MW}}
\end{figure}

We now have to minimize the real part of~$\wt V_2(\uu)/\wt \t^{\,2}$ as~$|\wt \t| \to 0$.
Since the~$u_{ij}$-independent piece and the real positive overall multiplicative constants 
are irrelevant in finding the dominant holonomy configurations, our problem boils down 
to minimizing the potential 
\begin{equation}
    \begin{split}
    V^Q(u_{ij};\arg\wt{\tau},\xi)\=-\mathrm{sign}(\arg\wt{\tau}-\frac{\pi}{2})\,
    \left(\overline{B}_3(m\xi+ m u_{ij})+\overline{B}_3(m\xi- m u_{ij})\right) \,,
    \label{eq:QhPot}
    \end{split}
\end{equation}
which is analogous to the \emph{pairwise holonomy potential} in~\cite{ArabiArdehali:2019tdm}. 
As in that work, we first consider the \emph{qualitative} behavior of $V^Q$. 
We assume $\mathrm{arg}(\wt{\tau})-\frac{\pi}{2}>0$, and comment below on what 
happens for the opposite sign.
We find that the potential is (see Figure~\ref{fig:MW}) 
\begin{equation*}
\begin{split}
    \text{M-shaped for }&0<\{m\xi\}<1/2 \,,\\
\text{W-shaped for }&1/2<\{m\xi\}<1 \,.
\end{split}
\end{equation*}
We also see from Equation~(\ref{defxi}) that we have~$\{m\xi\}\in\{0,\frac{1}{3},\frac{2}{3}\}$.

\subsubsection*{The $\rO(1/\wt \t^2)$ exponent}

Let us now assume $m,n$ are chosen such that~$\{m\xi\}=\{\frac{-m n_0-2n}{3}\}=\frac{1}{3}$, 
so we are in the M-region with the dominant holonomy configurations corresponding to~$\{m u_{ij}\}=0$. 
Although this is analogous to the~\emph{1-center phase} in~\cite{ArabiArdehali:2019orz}, 
as mentioned around~\eqref{eq:nontrivialSectors} 
here in fact $u_{ij}$ can be any integer multiple of~$\frac{1}{m}$. All these saddles contribute 
equally to the~$\rO(1/\wt\t^2)$ exponent though, and hence the preceding analysis 
around~$\uu=0$ gives the correct leading asymptotics of the index, which up to~$\rO(1/\wt \t)$ 
corrections in the exponent reads
\begin{equation}
\begin{split}
     &\exp \Bigl(-\frac{\pi \i}{m \, \wt \t^{\, 2}}(N^2-1) \overline{B}_3(m\xi) \Bigr) \=
        \exp\Bigl(-\frac{\i\pi }{27 m \, \wt \t^{\, 2}}(N^2-1)  \Bigr) \,,        
\label{eq:n0IndexRationalAsy1} \\ 
&\qquad \text{for} \quad \mathrm{arg} (\wt \t)>\frac{\pi}{2} \,, \quad \{m\xi\}\=\{\frac{-m n_0-2n}{3}\} \=\frac{1}{3} \,. 
\end{split}
\end{equation}
For $\mathrm{arg} (\wt \t)-\frac{\pi}{2}<0$, the M- and W-regions switch places. 
So in order to have $u_{ij}=0$ as the dominant saddle we must assume $m,n$ are such 
that $ \{m\xi\}=\{\frac{-m n_0-2n}{3}\}=\frac{2}{3}$. In this case we 
have~$\overline{B}_3(2/3) = - \overline{B}_3(1/3) =1/27$, which leads to 
\begin{equation}
\begin{split}
     &\exp \Bigl(-\frac{\pi \i}{m \, \wt \t^{\, 2}}(N^2-1) \overline{B}_3(m\xi) \Bigr) \=
        \exp\Bigl(\frac{\i\pi }{27 m \, \wt \t^{\, 2}}(N^2-1)  \Bigr) \,,        
\label{eq:n0IndexRationalAsy2} \\ 
&\qquad \text{for} \quad \mathrm{arg} (\wt \t)<\frac{\pi}{2} \,, \quad \{m\xi\}\=\{\frac{-m n_0-2n}{3}\} \=\frac{2}{3} \,. 
\end{split}
\end{equation}

In the remaining case where~$\{m\xi\}=\{\frac{-m n_0-2n}{3}\}=0$, we have~$\wt V_2 (\uu;\xi)=0$ 
and hence no~$\rO(\frac{1}{\wt \tau^2})$ exponent. As we discuss momentarily there is 
no~$\rO(\frac{1}{\wt \tau})$ exponent in this case either. There are thus~$\mathrm{rk}(G)$ 
flat directions in the moduli space, leading to a~$(1/\wt{\t})^{\mathrm{rk}(G)}$ growth for the index, 
as in the~$n_0=0$ and~$\tau$ pure imaginary case studied in~\cite{Ardehali:2015bla}.

\subsubsection*{The $\rO(1/\wt \t)$ exponent}

The~$\rO(1/\wt \t)$ exponent comes from $\wt V_1/m\wt\t$. Although the expression 
for~$\wt V_1$ in~\eqref{eq:simplifiedVtildes} was obtained near~$\uu=0$, the~$\rO(1/\wt \t)$ 
exponent is correctly captured by~\eqref{eq:gammaRationalEstWithR}, which implies 
that~\eqref{eq:simplifiedVtildes} remains correct near the nontrivial 
saddles with~$u_{ij}\in\frac{1}{m}\mathbb{Z}$ as well.
So we can specialize~$\wt V_1$ in~\eqref{eq:simplifiedVtildes} to the SU($N$) $\mathcal{N}=4$ theory
and obtain 
\begin{equation}
    \exp \Bigl(- \frac{\pi \i}{m \wt \t} (N^2-1) \, \bigl(-\overline{B}_2(m\xi)+\tfrac{1}{6} \bigr) \Bigr) \,.\label{eq:N=4at1/tau}
\end{equation}
In this case we have that~$\overline{B}_2(2/3) = +\overline{B}_2(1/3) = - 1/18$, which leads to  
\begin{equation}
    \exp \biggl(-\frac{2\pi \i \,}{9}\frac{(N^2-1)}{m \wt \t}\biggr) \,,
\end{equation}
for~$\mathrm{arg} (\wt \t)>\frac{\pi}{2}$  as well as~$\mathrm{arg} (\wt \t)<\frac{\pi}{2}$. 
Note that since~$\overline{B}_2(0) = 1/6$, we see from~\eqref{eq:N=4at1/tau} that there is
no~$\rO(1/\wt\t)$ exponent for~$\{m\xi\}=0$, as alluded to above.

\subsubsection*{The Chern-Simons coupling}

Specializing the Chern-Simons coupling \eqref{eq:CScouplingRat} to SU($N$) $\mathcal{N}=4$ theory we find
\be
k_{ij}=-\wt\eta \, N \, \delta_{ij},
\ee
with
\be
\wt\eta\defeq 6\overline{B}_1(m\xi)\= 6\overline{B}_1\bigl(\frac{-mn_0-2n}{3}\bigr) \,.\\
\ee
\vspace{.5cm}

We emphasize that all the topologically nontrivial sectors necessary for agreement with 
an~$S^3/\mathbb{Z}_m$ partition function are present in our analysis, but we leave the 
investigation of their explicit contributions to future work.

\subsubsection*{The $\rO(\wt \t)$ exponent}

The linear (in $\wt{\tau}$) exponent can be read from~\eqref{eq:In0semi-simpleAsyRat1} to 
be~$-2\pi i\wt\tau E_{\mathrm{susy}}/m$. Note again that while~\eqref{eq:In0semi-simpleAsyRat1} 
was derived near~$\uu=0$, as the estimate~\eqref{eq:denomEstRational0.75} shows the~$\rO(\wt \t)$ 
exponent remains valid near~$\uu\in\frac{\mathbb{Z}}{m}$ as well (at least for~$n=1$, and 
we expect more generally as well). Since for SU($N$) 
$\mathcal{N}=4$ theory $E_{\mathrm{susy}}= \frac{4}{27}\big(N^2-1\big)$, we have the $\mathcal{O}(\wt{\tau})$ 
exponent as in
\begin{equation}
    \exp \biggl(-\frac{8\pi \i}{27m}(N^2-1) \wt{\tau} \biggr) \,.
\end{equation}

\subsection*{Summary: the small-$\wt\t$ asymptotics for $\mathcal{N}=4$ SYM}

We can summarize the asymptotics of the SU($N$) $\mathcal{N}=4$ SYM index analyzed above as follows
\begin{equation}
    \CI_N(\t;n_0)  \; \simeq \; N \, \wt {C}_N(n_0,m,n)\,  
    \exp \biggl(-\frac{\i\pi}{m \,\wt\tau^2} \, (N^2-1) \Bigl(\frac{-\wt \eta+2\wt\tau}{3} \Bigr)^3 \biggr) \;
Z^{\text{CS}}_{S^3/\mathbb{Z}_m}(k)  \,,
  \label{eq:N=4indexAsyRational}
\end{equation}
for $\tau$ near any rational point~$-n/m$, with  
\be
\wt \t \= m\t+n \,, \qquad \wt \eta \= 6\overline{B}_1(\frac{-mn_0-2n}{3}) \= -\mathrm{sign}(\mathrm{arg}(\wt \t)-\tfrac{\pi}{2}) \,,
\qquad k \=- \wt \eta \, N  \,,
\ee
and with~$\wt {C}_N(n_0,m,n)$ an overall constant. Note that we have used~$\wt{\eta}^3=\wt{\eta}=\pm1$ 
to simplify the final expression. 
Also, by completing the cube inside the exponent we have introduced an~$\mathcal{O}(\wt{\tau}^0)$ 
factor at the cost of redefining~$\wt{C}_N(n_0,m,n)$.

We have only demonstrated that there is a contribution to~$Z^{\text{CS}}_{S^3/\mathbb{Z}_m}(k)$ from 
near~$\uu=0$ that coincides with the topologically trivial sector of the~$S^3/\mathbb{Z}_m$ 
partition function of Chern-Simons theory with coupling~$k$. As mentioned below~\eqref{eq:nontrivialSectors} 
we expect that summing over the contributions from neighborhoods of the non-trivial 
configurations~$u_j=m_j/m$ would lead to the complete orbifold partition function.

We can include the contribution of a decoupled $U$(1) $\mathcal{N}=4$ multiplet in a straightforward manner. 
This effectively changes the dimension of the group in the exponent to~$N^2$, introduces a prefactor~$1/\wt \t$,
and change the constant from~$\wt{C}_N(n_0,m,n)$ to a new constant~$\wt{C'}_N(n_0,m,n)$, so that we have
\begin{equation}
    \CI^{U(N)}(\t;n_0)  \; \simeq \; \frac{N}{\i \wt \t} \, \wt{C'}_N(n_0,m,n) \, 
    \exp \biggl(-\frac{\i\pi}{m \,\wt\tau^2} \, N^2 \Bigl(\frac{-\wt \eta+2\wt\tau}{3} \Bigr)^3 \biggr) \;
Z^{\text{CS}}_{S^3/\mathbb{Z}_m}(k)  \,.
  \label{eq:N=4indexAsyRationalU(N)}
\end{equation}
We see that the background (and the SUSY Casimir) piece in~\eqref{eq:N=4indexAsyRationalU(N)} matches the effective 
action~\eqref{actionEll} and, in addition, we have a dynamical Chern-Simons term.
In the following section we explain both these pieces from the point of view of 3d $\mathcal{N}=2$ field theory.

\subsection{$C$-center phases}\label{sec:Ccenter}

Focussing on SU($N$) $\mathcal{N}=4$ theory, we now move on to studying the $\wt\t\to0$ limit 
of the index in the \emph{W region}, which as shown in Figure~\ref{fig:MW} for 
$\mathrm{arg}\wt{\tau}>\pi/2$ corresponds to $1/2<\{m\xi\}<1$. As before we assume 
$\mathrm{arg}\wt\t$ is in compact domains avoiding integer multiples of $\pi/2$ as $|\wt\t|\to0$.

Recall from (\ref{defxi}) that only the values $\{m\xi\}=0,1/3,2/3$ are realized in our problem. 
But to highlight the parallels with the analysis of partially-deconfined phases in the W regions 
of the (flavored) 4d $\mathcal{N}=4$ index in \cite{ArabiArdehali:2019orz}, we will study the 
phase structure for arbitrary $\{m\xi\}\in(\frac{1}{2},1)$ below, and only at the end specialize 
our result to the single ``physical'' point $\{m\xi\}=2/3$ in that interval.

Asymptotic analysis of the index for arbitrary $\{m\xi\}\in(\frac{1}{2},1)$ is difficult for general $N$, 
because finding the dominant holonomy configurations is not possible analytically in the W regions. 
Analogously to \cite{ArabiArdehali:2019orz} we consider now the large-$N$ limit (on top of 
the $\wt\t\to0$ limit), and conjecture that the $C$-center phases suffice for extremizing the 
potential in the W region. Also, similarly to \cite{ArabiArdehali:2019orz} we consider only the 
leading (here $\rO(1/\wt\t^2)$) exponent of the index in the W region.

A $C$-center holonomy configuration consists of $C$ packs of $N/C$ holonomies uniformly 
distributed on the circle such that the SU($N$) constraint is satisfied. While at finite $N$ it is 
possible to have such configurations only for $C$ a divisor of $N$, in the large-$N$ limit any 
integer $C\ge1$ provides an acceptable $C$-center configuration \cite{ArabiArdehali:2019orz}. 
For such a distribution the ``on-shell'' value of the potential $\wt V_2$ in (\ref{eq:QhDef}) becomes
\begin{equation}
   \wt V_2^{(C)}=\i\pi\Bigl((N-1)\overline{B}_3(m\xi)+\frac{N}{d}\frac{d(d-1)}{2}\,
   2\overline{B}_3(m\xi)+d^2 \sum_{J=1}^{C-1}J\big(\overline{B}_3(m\xi+m\frac{J}{C})
   +\overline{B}_3(m\xi-m\frac{J}{C})\big)\Bigr),
\end{equation}
where $d:=N/C$. The second term above is the contribution from pairs in the same pack, and the 
third term is from pairs with each end on a different pack. To simplify the above expression further, 
we use the following identity which can be proven from (\ref{eq:Raabe}) and (\ref{eq:remarkableId}):
\begin{equation}
    \sum_{J=1}^{C-1}J\big(\overline{B}_3(\Delta+m\frac{J}{C})+\overline{B}_3(\Delta-m\frac{J}{C})\big)
    =\frac{g^2\overline{B}_3(C'\Delta)}{C'}-C\overline{B}_3(\Delta),
\end{equation}
where $g:=\mathrm{gcd}(m,C)$ and $C':=C/g$. Keeping only the $O(N^2)$ terms we hence end up with
\begin{equation}
    \wt V_2^{(C)}=\i\pi N^2\,\frac{\overline{B}_3(C'm\xi)}{C'^3}.
\end{equation}
Since the leading asymptotics of the index is given as $\exp(-\wt V_2/\wt\t^2)$, we then find the 
analog of the main result of \cite{ArabiArdehali:2019orz} (Equation~(3.19) of that work) for our case to be
\begin{equation}
    \mathcal{I}_{N\to\infty}\xrightarrow{\wt\t\to0}\sum_{C=1}^{\infty}
    \exp\left(-\frac{i\pi N^2}{m\wt{\tau}^{\,2}}\,\frac{\overline{B}_3(C'm\xi)}{C'^3}\right),
    \label{eq:doubleScalingConjecture}
\end{equation}
with $m\xi=-\frac{mn_0+2n}{3}$ as before.

\begin{figure}[h]
\centering
    \includegraphics[scale=.45]{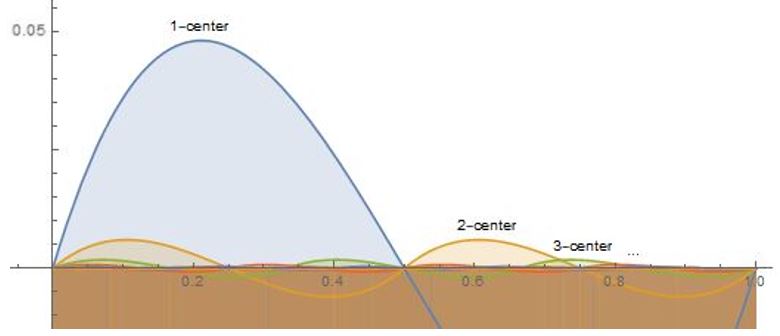}
\caption{The functions $C'^{-3}\overline{B}_3(C'\Delta)$ with
$C'=1,\cdots,13$, for $0\le\Delta\le1$. For $0<\Delta<1/2$ the blue curve corresponding to the 
fully-deconfined phase takes over. The take-over of the
orange curve signifies the partially-deconfined 2-center phase in the corresponding 
region ($1/2<\Delta\lesssim .72$),
and so on. \label{fig:singleDelta}}
\end{figure}

The competition between various terms in (\ref{eq:doubleScalingConjecture}) can be visualized by 
comparing the exponents as in Figure~\ref{fig:singleDelta}, which shows the range of $\Delta:=\{m\xi\}$ 
for which a given phase dominates when $\mathrm{arg}\wt\t-\pi/2>0$. The figure implies that for the 
``physical'' values $\{m\xi\}=1/3,2/3$, the index is respectively in the 1-center, and 2-center phase 
when $\mathrm{arg}\wt\t-\pi/2>0$, and vice versa for $\mathrm{arg}\wt\t-\pi/2<0$. As mentioned above, 
for $\{m\xi\}=0$ the index is in a confined phase and does not yield exponential $O(N^2)$ growth. 
Therefore up to an $o(N^2/\wt{\tau}^{2})$ error in the exponents we have the following simplification of 
(\ref{eq:doubleScalingConjecture}) by restricting to $C'=1,2$:
\begin{equation}
    \mathcal{I}_{N\to\infty}\xrightarrow{\wt\t\to0}e^{-\frac{i\pi N^2}{m\wt{\tau}^2}\,
    \overline{B}_3(m\xi)}+e^{-\frac{i\pi N^2}{m\wt{\tau}^2}\,\frac{\overline{B}_3(2m\xi)}{8}}.
    \label{eq:doubleScalingConjecture2}
\end{equation}
This is the analog of Conjecture~1 in \cite{ArabiArdehali:2019orz}.

Since $\overline{B}_3(2/3)=-\overline{B}_3(1/3)$, we see from~\eqref{eq:doubleScalingConjecture2} 
that the action of the 2-center saddle has the opposite sign and is smaller in absolute value by a 
factor of 8 compared to that of the 1-center saddle.

\section{Asymptotics of the 4d index from 3d field theory \label{sec:4dto3d}}

In this section we consider the dimensional reduction of the four-dimensional~$\CN=1$ 
gauge theory on a Hopf surface. This surface is topologically~$S^3 \times S^1$ and  
we reduce along the~$S^1$ fiber.  
The dimensionally reduced theory describes a three-dimensional dynamical gauge 
supermultiplet coupled to background three-dimensional supergravity on~$S^3$. 
The Wilsonian effective action of the gauge multiplet can be calculated by integrating 
out the tower of massive Kaluza-Klein modes, and the resulting theory is described by a 
functional integral over the gauge multiplet fields with this effective action. 
We find that the functional integral of the three-dimensional theory 
can be written as a perturbative expansion in~$\t$. 
The singular terms in the expansion behave as~$\rO(1/\t^2)$ and~$\rO(1/\t)$, and are captured 
by three-dimensional effective field theory. 
In particular, these terms are independent of the dynamical fields, and are completely 
accounted by the (supersymmetrized) Chern-Simons couplings of the background supergravity. 
The result agrees with the corresponding singular terms in the microscopic 
expansion~\eqref{eq:almostThere},~\eqref{eq:Zback}.

The all-order asymptotic formula from the microscopic index includes, in addition to these singular terms, 
constant and linear terms in~$\tau$. Using a localization argument we show that 
the constant term in~$\t$, besides a background part, has a dynamical piece captured by the 
integral over the fluctuations of the dynamical fields in three-dimensional path integral, which is 
essentially the partition function of~$\CN=2$ supersymmetric CS theory at level~$\pm N$. 
Finally, the linear term in the microscopic formula is precisely the supersymmetric Casimir energy 
which is needed to translate between the microscopic Hamiltonian index and the macroscopic 
functional integral.\footnote{The supersymmetric Casimir energy that appears in our asymptotic 
formulas is the one given in~\cite{Assel:2015nca}. Note in particular that (unlike 
in~\cite{Cabo-Bizet:2018ehj}) this is independent of~$n_0$. 
We can understand this in the path-integral picture by appealing to the result in Section~4 of~\cite{ArabiArdehali:2019tdm} 
(based on the regularization method of \cite{Ardehali:2015hya}) which demonstrated that the 
supersymmetric Casimir energy is independent of flavor fugacities when they are on the unit circle, 
and by noting that~$\rme^{2\pi \i(-n_0 r_I/2)}$ is effectively a flavor fugacity in our problem.}
In this manner the full asymptotic formula for the four-dimensional index is explained 
by three-dimensional physics. 
The fact that the asymptotic formula does not contain any higher order terms in~$\t$ 
implies a non-renormalization theorem, namely 
that there are no corrections to the three-dimensional effective action at any polynomial order in~$\t$. 
We leave the explanation of this interesting point to future work.  
Finally, we show that corresponding statements also hold near rational points 
when~$\t \to -n/m$. Here we present evidence that the relevant three-dimensional manifold is a~$\mathbb{Z}_m$ orbifold of~$S^3$ and the results agree with the microscopic asymptotic expansion given 
in~\eqref{eq:N=4indexAsyRational}.

\vskip 0.4cm

We begin by recalling the functional integral definition of the~$\CN=1$ superconformal index on~$S^3 \times S^1$. 
In the Hamiltonian trace definition~\eqref{defindex} we have two chemical potentials that couple to linear 
combinations of the two angular momenta~$J_1, J_2$ on~$S^3$ and the~$U(1)$ R-charge~$Q$. 
This is equal to the supersymmetric functional integral of the theory on~$S^3 \times S^1$ with twisted boundary 
conditions on the fields as we go around the~$S^1$.
Equivalently, one can explicitly introduce a background gauge field (for the R charge) and background 
off-diagonal terms in the metric (for the angular momenta) in a manner, so as to preserve supersymmetry.  
As explained in \cite{Festuccia:2011ws}, such background configurations can be obtained as solutions to 
the condition of vanishing gravitino variations of off-shell supergravity (and then taking a rigid limit so as to decouple the 
fluctuations of gravity).

The relevant background configuration for the calculation of the 4d superconformal index for 
complex~$\tau$ and nonzero $n_0$ was studied
in~\cite{Cabo-Bizet:2018ehj} in the context of 4d new minimal supergravity~\cite{Sohnius:1981tp, Sohnius:1982fw}.
Recall that the bosonic fields of new minimal supergravity are the metric, a gauge field~$A^\text{nm}$, 
and another vector field~$V^\text{nm}$ which is covariantly conserved. 
The background configuration~\cite{Cabo-Bizet:2018ehj} preserving the supercharges~$(\CQ, \CQb)$ 
is\footnote{A real metric corresponds to pure imaginary~$\Omega_i$. 
General complex $\Omega_i$ correspond to analytic continuation in the background metric.} 
\be \label{4dbackgnd}
\begin{split}
\dd s_4^2 & \=  \dd t_E^2 + \dd \theta^2 + \sin^2 \theta \, 
\bigl(\dd \phi_1 -\i \,\Omega_1 \, \dd t_E \bigr)^2 + \cos^2 \theta \, \bigl(\dd \phi_2 -\i \, \Omega_2 \, \dd t_E \bigr)^2  \,,\\
A^\text{nm} & \= \i\, \Bigl(\Phi - \frac{3}{2} \Bigl) \dd t_E \,, \qquad V^\text{nm}  \=  -\i \,  \dd t_E \, .
\end{split}
\ee
Here~$\theta \in [0, \pi/2]$, the angles~$\phi_1$, $\phi_2$ are $2\pi$-periodic, and 
the Euclidean time coordinate has the independent periodicity 
condition\footnote{In~\cite{Cabo-Bizet:2018ehj} the parameter~$\rr$ was called~$\b$.} 
\be
t_E \sim t_E + \rr  \,.
\ee
This configuration admits the following Killing spinor which is identified with~$\CQ$, 
\be \label{KSsol}
\varepsilon \=
\left(
\begin{array}{c}
\rme^{\i \, z \,t_E } \\
0 \\
0 \\
\rme^{- \i \, z \, t_E } \\
\end{array}
\right) \,, \qquad z \= \frac{\pi n_0}{\rr}  \,.
\ee

The twist parameters~$\Omega_i$, $\Phi$ are related to the chemical potentials~$\s$,~$\t$ in the index as 
follows\footnote{Here~$(\Omega_1^*,\Omega_2^*, \Phi^*) = (1,1,\frac32)$ are the values of the potentials
on the supersymmetric BH solution.}
\be
\Omega_i \= 1 + \frac{\omega_i}{\rr} \,, \qquad \Phi \= \frac32 +  
\frac{1}{\rr} \Bigl(\frac{\omega_1+\omega_2}{2} - \pi \i \, n_0 \Bigr) \,,
\label{Omomrel}
\ee 
with 
\be
\omega_1 \= 2 \pi \i \,\s \,, \qquad \omega_2 \= 2 \pi \i \, \t \,.
\ee
In this section for ease of presentation we focus on the case with~$\Omega_1=\Omega_2=\Omega,$ 
which implies~$\sigma=\tau=\frac{\i\rr}{2\pi}(1-\Omega)$. 
The partition function on the above background is related to the index~$\CI(\sigma-n_0,\tau)$, 
which for $\sigma=\tau$ coincides with the index~$\CI(\tau;n_0)$ in~\eqref{eq:n0Index}. 
In Appendix~\ref{sec:diffomegas} we comment on the more general case 
with~$\O_1\neq\O_2$ and hence $\sigma\neq\tau$.

The four-dimensional supersymmetric partition function of the theory corresponding to the Hamiltonian 
index~\eqref{defindex} can then be expressed as a functional integral 
of the gauge theory with 4d~$\CN=1$ chiral and vector multiplets on the  
background~\eqref{4dbackgnd}.\footnote{More precisely the Hamiltonian index equals the functional 
integral for the supersymmetric partition function up to the supersymmetric Casimir energy 
factor~\cite{Ardehali:2015hya,Assel:2015nca}.}
As discussed in~\cite{Cabo-Bizet:2018ehj}, this functional integral 
localizes to an integral over flat connections of the gauge field on the KK circle, 
\be \label{AYval}
\oint A^i \= 2 \pi \, u_i \,.
\ee
The Wilson loop~\eqref{AYval} maps to the scalar in the three-dimensional vector multiplet 
in the KK reduction. 
We now proceed to derive an expression for the supersymmetric partition function of 
the three-dimensional gauge theory.

\subsection{Dimensional reduction to three dimensions \label{sec:dimred}}

We first consider the reduction of the above four-dimensional background as a 
configuration in three-dimensional supergravity. 
In three dimensions we use the off-shell supergravity 
formalism~\cite{Kuzenko:2011rd, Kuzenko:2011xg,  Kuzenko:2012bc, Kuzenko:2013uya}, 
and follow the treatment~\cite{Closset:2012vp, Closset:2012vg, Closset:2012ru, Assel:2014paa} 
for the reduction from four to three dimensions.  The bosonic fields in the off-shell three-dimensional 
supergravity are the metric, the KK gauge field (the graviphoton) written as a one-form~$c$, a 
two-form~$B$, and the R-symmetry gauge field one-form~$\CA^R$. 
The equations are often presented in terms of the dual one-form~$v=-\i *\dd c$
and the dual scalar~$H = \i * \dd B$.

We begin by writing the background in~\eqref{4dbackgnd} as a Kaluza-Klein (KK) compactification 
to three dimensions, i.e.~a circle fibration on a 3-manifold~$\CM_3$.
We define the rescaled~$S^1$ coordinate 
\be \label{X4period}
Y \= \sqrt{1 - \Omega^2} \; t_E \,,
\ee
which obeys the periodicity condition 
\be
Y \sim Y + 2 \pi R\,, \qquad  R \= \frac{\rr}{2 \pi}  \sqrt{1-\O^2} \,.
\label{Rgamrel}
\ee

Writing the metric~\eqref{4dbackgnd} in the KK form,
\be  \label{4dmetricKKform}
\dd  s_4^2  \=  \dd s_3^2 + (\dd Y + c)^2 \,,
\ee 
we find that the graviphoton field is 
\be \label{3dKKc}
c \= c_\mu \, \dd x^\mu   
 \= -\i \frac{\O}{\sqrt{1-\O^2}} \, \bigl( \sin^2 \theta \,\dd \phi_1  + \cos^2 \theta \,\dd \phi_2 \bigr) \,,
\ee
and the metric on the 3-manifold~$\CM_3$ is 
\be \label{3dmetric1}
\dd s_3^2  \= g_{\mu\nu} \, \dd x^\mu \, \dd x^\nu 
\=  \dd \theta^2 + \sin^2 \theta \,  \dd \phi_1^2 + \cos^2 \theta \, \dd \phi_2^2  - c^2 \,.
\ee
The three-dimensional metric obeys 
\be \label{sqrtg}
\sqrt{g} \= \frac{\sin 2 \theta}{2 \sqrt{1-\O^2}} \,.
\ee
We see that we effectively have a KK reduction on a circle of radius~$R$. 

\vskip 0.4cm

In order to study the effective theory in three dimensions, we consider the limit~$R \to 0$.
From the relation~\eqref{Rgamrel} we see that this is implemented by taking the original circle size~$\rr \to 0$. 
Our eventual interest is in the limit~$\t \to 0$. The question is how to correlate these two limits of~$\rr$ and~$\t$.  
If we take~$\rr \to 0$ first, then we see from the relation~\eqref{Omomrel} that~$\Omega \to \infty$ 
and from~\eqref{sqrtg} that~$\CM_3$ shrinks to zero size. 
Although the local Lagrangian involves background fields and terms such as the Ricci scalar which diverge in this limit, 
the three-dimensional effective action turns out to be finite.  
We can understand this in a cleaner manner as follows. 
We first scale~$\t$ and~$\rr$ to zero at the same rate keeping~$\O$ finite and fixed, 
i.e.~take~$\rr = \ve \t$ with fixed~$\ve = 2 \pi \i /(\O -1)$, and only take~$\ve\to0$ at the end of all calculations. 
In particular, the three-dimensional calculations
are all performed at finite~$\ve$, i.e.~on smooth backgrounds. 
The action turns out to have two pieces, one of which stays finite and the other vanishing in the limit~$\ve \to 0$,
and, in particular, there are no diverging terms in this limit. 
Thus we can safely take the limit~$\O \to \infty$ at the end of calculations. 
In this limit we have that~$R \to \t$, so that the effective field theory answers are effectively written as 
a perturbative series in~$\t$.

\vskip 0.4cm

In the treatment of three-dimensional background supergravity we need the Hodge dual of the graviphoton,
\be \label{defv}
v \= v_\mu \, \dd x^\mu \= - \i * \dd c \,,
\ee
whose value in the above background is
\be
v \= \frac{2 \, \O}{1-\O^2} \, \bigl( \sin^2 \theta \,\dd \phi_1  + \cos^2 \theta \,\dd \phi_2 \bigr) \,,
\ee
so that~$v^\mu = 2 \,\O(1,1,0)$. 
The associated Chern-Simons action is 
\be
S^\text{CS} (c)  \= \int_{\CM_3} \, c \wedge \dd c  \= \i \int_{\CM_3} \dd^3 x \,  \sqrt{g} \, v^\mu \, c_\mu\,. 
\ee

\vskip 0.4cm

The identification between the four-dimensional and the three-dimensional gauge fields 
is made by comparing the respective Killing spinor equations. 
As shown in~\cite{Closset:2012ru, Assel:2014paa}, one has\footnote{In~\cite{Closset:2012ru} it is 
assumed that $V^{\text{nm}}_Y=A^{\text{nm}}_Y$, which is not satisfied in our background. 
Therefore we follow more closely the treatment of~\cite{Assel:2014paa}.} 
\be
\frac12 v_\mu \= V^\text{nm}_\mu - V^\text{nm}_Y c_\mu \,, \qquad H \= V^\text{nm}_Y \,,
\qquad \CA^R_\mu \= A^\text{nm}_\mu - A^\text{nm}_Y c_\mu  + \frac12 v_\mu \,.
\ee
The background gauge fields in~\eqref{4dbackgnd} are given by
\be \label{AYVYval}
A^\text{nm}
 \=  \Bigl( -\frac{\t}{R}+ \frac{n_0}{2R} \Bigr) \, \dd Y \,, 
\qquad V^\text{nm}  \=  -\frac{\i}{\sqrt{1-\O^2}} \,  \dd Y \,,
\ee 
so that the auxiliary fields in the background supergravity multiplet are 
\be \label{vHvalues}
v_\mu \=  - 2 \, V^\text{nm}_Y \, c_\mu  \,, \qquad  
H \= V^\text{nm}_Y \,, \qquad 
\CA^R_\mu \= - (A^\text{nm}_Y + V^\text{nm}_Y) \, c_\mu \,.
\ee
(The above equation for~$v_\mu$ is consistent with Equations~\eqref{3dKKc},~\eqref{defv}.)

\vskip 0.4cm

We now discuss the Kaluza-Klein reduction of the dynamical gauge multiplet. 
The~$\CN=1$ gauge multiplet in four dimensions reduces to an~$\CN=2$
gauge multiplet in three dimensions, whose bosonic field content is a vector~$\CA_\mu$, 
a scalar~$\s$, and  the auxiliary~$\CD$ field. 
These are related to the four-dimensional fields as follows,
\be\label{eq:susy3dvec}
\sigma^i \= A^i_Y \,, \qquad \CA^i_\mu \= A^i_\mu -  A^i_Y \, c_\mu \,, \qquad \CD^i \= D^i -  A^i_Y \, H \,,
\ee
and the three-dimensional fermions are the reduction of the corresponding four-dimensional fermions.
As discussed above, the theory localizes on the BPS configurations given by 
\be 
A^i \= \frac{u_i}{R}  \, dY \,, \qquad D^i \= 0 \,, 
\label{AYval}
\ee
with vanishing values of all other fields in the off-shell gauge and chiral multiplets. 
In the three-dimensional theory the non-zero fields on the BPS locus are 
\be
\sigma^i \= \frac{u_i}{R} \,, \qquad 
\CA^i_\mu \=   - \frac{u_i}{R} \, c_\mu \,, \qquad \CD^i \=  -  \frac{u_i}{R}  \, H \,.
\label{AYval3d}
\ee

\subsection{Effective action and functional integral of the three-dimensional theory \label{sec:3deffact}}

We now turn to the calculation of the partition function of the three-dimensional supersymmetric theory
that we just discussed. 
Our strategy is to first calculate the three-dimensional Wilsonian effective action of~$u_i$, and then use 
this to calculate the three-dimensional partition function. 
The tree-level action (coming from a mode expansion of the four-dimensional theory) 
consists of matter-coupled super Yang-Mills theory.
The full quantum effective action of the three-dimensional theory is obtained by integrating out the 
tower of massive KK modes on the circle. In order to calculate this action, we draw from known results 
in the effective field theory in three dimensions. 

The effective field theory on backgrounds of the type~$\CM_3 \times S^1_R$ was studied in a  
general context in~\cite{Banerjee:2012iz,Jensen:2012jh},
and in the special context of supersymmetry in~\cite{DiPietro:2014bca, DiPietro:2016ond}.  
The resulting three-dimensional action begins with a term proportional to~$1/R^2$,
and continues as a perturbation expansion as the radius~$R \to 0$. 
At each order in~$R$ one has 
a combination of three-dimensional actions of the background and the dynamical fields, 
which are all related by supersymmetry to a certain Chern-Simons term.
The Chern-Simons terms are of the form~$\int_{\CM_3} A_x \wedge \mathrm{d}A_y$, where~$A_x$
and~$A_y$ represent the various gauge fields. As discussed in the previous subsection, 
these are the dynamical gauge field, the background graviphoton, the background R-gauge field, 
and the spin connection. We follow, and review in Appendix~\ref{sec:CSactions}, 
the treatment of~\cite{DiPietro:2016ond} for the supersymmetrized  Chern-Simons action of all 
the background and the dynamical gauge fields up to~$O(R^0)$. The full effective action also 
includes RR and gravitational supersymmetrized CS terms discussed in \cite{Closset:2018ghr}, 
which turn out to be crucial for our purposes.

It follows from the above discussion that the overall coefficient at each order in~$R$ 
can be fixed by calculating the coefficient of the Chern-Simons terms themselves. 
These coefficients, in turn, can be obtained by integrating out all the fermions coupling 
to the corresponding gauge fields. The resulting induced Chern-Simons coefficient is one-loop exact. 
Thus the strategy is to integrate out the fermions in each KK mode, write the 
resulting Chern-Simons action, and sum over all the fermions in the theory. 
The KK momenta of the fermions take the values ~$p_Y = k_Y/R$, with $k_Y =n + \frac{n_0}{2}$, $n \in \IZ$.
The shift~$n_0/2$ appears because of the gauge fields in the background~\eqref{4dbackgnd}. 
(Recall, for example, that the four-dimensional Killing spinor~\eqref{KSsol} has momentum~$n_0/2$.)

The result for the complete action obtained by integrating out a fermion~$\tf$ of 
R-charge~$r_\tf$ and transforming in a representation of weight~$\rho_\tf$ under the gauge group 
is given in Appendix~\ref{sec:CSactions} and take the following form,
\be
\delta S^\tf_\text{1-loop} \= \wt{S}^\tf_\text{g-g} + 2 \, \wt{S}^\tf_\text{g-R} + S^\tf_\text{R-R} + S^\tf_\text{grav} \,.
\label{Sftot1}
\ee
The terms in~\eqref{Sftot1} depend on the real mass~$m_\tf$ (related to the central charge appearing 
in the three-dimensional algebra). 
The first two terms depend on the dynamical gauge field. 
On the configuration~\eqref{AYval3d} they take the following values, 
\be
\begin{split}
\wt{S}^\tf_\text{g-g} & \=-\i \pi \, \frac{ \mathrm{sgn}(m_\tf)}{8R^2} \,  
\bigl(\rho_{\tf}\cdot \uu - k_Y \bigr)^2\, A_{\mathcal{M}_3} \,, \\
2 \, \wt{S}^\tf_\text{g-R} & \=-\i \pi \,  \frac{\mathrm{sgn}(m_\tf)}{8R} \, 
2 \, r_\tf  \, \bigl(\rho_{\tf}\cdot \uu - k_Y \bigr) \; L_{\mathcal{M}_3} \,,
\end{split}
\label{S1S2fin}
\ee
where~$A_{\mathcal{M}_3}$ and $L_{\mathcal{M}_3}$ are functions
of the three-dimensional background given in~\eqref{defALl}.
The last two terms in~\eqref{Sftot1} do not depend on the dynamical gauge field,
and given by 
\be
\begin{split}
    S^\tf_\text{R-R}&\=-\i \pi \, \frac{\mathrm{sgn}(m_\tf)}{8} \, \bigl(r_\tf^2-\frac{1}{6}\bigr) \, R_{\CM_3}  \,, \\
  S^\tf_\text{grav}&\=-\i \pi \, \frac{\mathrm{sgn}(m_\tf)}{192} \, G_{\CM_3} \,, 
\label{S:susy:explicit3}%
\end{split}
\ee
where~$R_{\CM_3}$ and~$G_{\CM_3}$ are functions
of the three-dimensional background given in~\eqref{eq:CSactions2}.

In Appendix~\ref{sec:CSactionvals} we calculate the values of these background  
actions.\footnote{We note that there is a subtlety with the gravitational CS term in~\eqref{eq:CSactions2},
concerning the dependence of the term on the frame~\cite{Witten:1988hf}. 
There should be a choice of frame which is consistent with the supersymmetry and the 4d to 3d reduction.
We do not work out the details of this issue in this paper, and instead rely on consistency 
with~\cite{Closset:2018ghr} where this term is obtained indirectly by considering integrating out chiral multiplets.
We thank Cyril Closset for a discussion on this point.}
As explained above, we perform the calculations keeping~$R$, $\O$ finite so that 
the three-dimensional physics is manifestly smooth. The result is that there is a smooth 
limit as~$\rr \to 0$ keeping fixed~$\t$. 
The limiting values of the actions are as follows,
\be
\begin{split}
&A_{\mathcal{M}_3}  \= - 4 \,, \qquad 
 L_{\mathcal{M}_3} \= - 4  \Bigl( 1 - \frac{n_0}{2R} \Bigr) \,, \\ 
& R_{\CM_3}  \= - 4 \, \Bigl( 1 -\frac{n_0}{2R} \Bigr)^2 \,, \qquad 
G_{\CM_3} \= - 16 + 4 \, R_{\CM_3} \,.
\end{split}
\label{ALvals}
\ee
Using these values, we obtain the total effective action of the fermion~$\tf$  to be
\be
\begin{split}
\delta S^\tf_\text{1-loop} 
& \= \i \pi \,  \frac{\mathrm{sgn}(m_\tf)}{2R^2}  \, 
\bigl(\rho_{\tf}\cdot \uu - k_Y - \tfrac{1}{2} n_0 \, r_\tf \bigr)^2 
+ \i \pi \,  \frac{\mathrm{sgn}(m_\tf)}{R}  \, r_\tf \, \bigl(\rho_{\tf}\cdot \uu - k_Y - \tfrac{1}{2} n_0 \, r_\tf \bigr) \\
& \qquad  + \i \pi \,  \frac{\mathrm{sgn}(m_\tf)}{2}  \, r_\tf^2 
\; - \; \i \pi \,  \frac{\mathrm{sgn}(m_\tf)}{12} \,.
\end{split}
\label{defSf}
\ee

\vskip 0.4cm

Now we turn to the sum over all the fermions in the theory. 
The value of the real mass is given in~\eqref{realmass} to be, 
as~$R \to 0$,\footnote{In fact the first three terms in~\eqref{defSf} sum up to
\be \i \frac{\pi}{2} \, \mathrm{sgn}(m_\tf)  
\Bigl( \frac{ \rho_{\tf}\cdot \uu - n - \tfrac{1}{2} n_0 \, r_I  + (r_I-1) R }{R}  \Bigr)^2 \,,
\ee
and, using~\eqref{eq:KKsumBbar} with~$x=  \rho_{\tf}\cdot \uu - n - \tfrac{1}{2} n_0 \, r_I  + (r_I-1) R$ 
to perform the sum over the KK modes, we obtain an effective potential which 
reproduces the chiral multiplet contributions in~\eqref{Veffans3d}. 
Essentially the same comment can be made in the microscopic analysis of Section~\ref{sec:SCI}.
} 
\be
m_{\tf,n} \= -\frac{1}{R} \bigl(\rho_\tf\cdot \uu - n - \tfrac12 n_0  \, (r_\tf+1)   \bigr) \,,
\ee
In order to obtain the full effective action we now have to sum over all the fermions.
For the chiral multiplets, this implies summing over all the weights in 
representations~$\rho_\tf\in\mathcal{R}_\tf$, as well as over all momenta labelled by $n\in\mathbb{Z}$. 
The summation over KK modes can be evaluated using
\begin{equation}
    \sum_{n\in\mathbb{Z}}\mathrm{sgn}(n+x)(n+x)^{j-1} \= -\frac{2}{j} \, \overline{B}_{j}(x) \,,
\label{eq:KKsumBbar}
\end{equation}
with~$x=\rho_\tf\cdot \uu - \tfrac12  n_0  \, r_I $, for $j=1,2,3$ (cf.~Section~4 
of~\cite{ArabiArdehali:2019tdm}).\footnote{Here, the~$\sgn$ function is interpreted as applying 
to~$m_\tf$ with~$R$ a real positive number (which therefore scales out of the formula so 
as to give~$\sgn(\rho_\tf\cdot \uu - \tfrac12  n_0  \, r_I)$). 
Note that in the subsequent formulas $R$ is taken to be complex.}
Here we have used the relation~$r_\tf = r_I-1$ between the R-charge of the fermion and that of the 
bottom component of the multiplet~$I$ to which the fermion belongs. 

\vskip 0.4cm

For the vector multiplet contribution the analysis is quite similar: there is a tower of massive KK gaugino 
modes that are integrated out. These generate CS actions whose supersymmetrization yields the 
vector multiplet contribution to $\delta S_{\text{1-loop}}$. In the present context there is an important difference 
with the chiral multiplet analysis however. Near $\uu=0$ there is a single gaugino mode in the tower that 
has real mass of order $\alpha\cdot\uu/R$, and is therefore considered a ``light'' mode 
for small enough~$|\alpha\cdot\uu|$. 
Therefore we do not integrate out this mode and, instead, keep it as a dynamical mode in 
the path integral of the three-dimensional theory.

More precisely, recall that the $n$th KK gaugino mode associated to a root~$\alpha$ of 
the gauge group has~$p_Y=(n+n_0)/R$ and hence a real mass~$(\alpha\cdot\uu -n-n_0)/R$. 
Therefore the mode corresponding to~$n=-n_0$ is light near~$\alpha\cdot\uu=0$. 
We now describe how removing this term from the sum over the KK tower modifies the result 
compared to the chiral multiplet computation. 
The vector multiplet contributions is a sum over roots~$\alpha$ that come in pairs~$\pm\alpha_+$, 
as a result of which they give vanishing contributions to the quadratic and constant terms in~$\uu$
in the action of a single KK mode.
We therefore focus on the contribution to the linear term in~$\uu \,,$ which is proportional to~$1/R$. 
The calculation is similar to the corresponding chiral multiplet calculation.  
Upon summing over all the KK modes, we obtain 
the vector multiplet contribution 
from a root~$\alpha$ to be \\
\be
-\frac{\pi \i}{R} \, \sum_{n\in\mathbb{Z}}{}^{'}\mathrm{sgn} \, (\alpha\cdot\uu -n-n_0)\, 
\bigl(\alpha\cdot\uu -n-n_0\bigr)\label{eq:sumPrimeVec} \,,
\ee
where the prime indicates that we are not including the light mode corresponding to $n=-n_0$. 
Upon adding and subtracting the $n=-n_0$ contribution, we obtain, using~\eqref{eq:KKsumBbar},  
\be
\frac{\i\pi}{R} \, \Bigl( \overline{B}_2(\alpha\cdot\uu)+ |\alpha\cdot\uu|  \Bigr)\,.
\ee
Now, since we are interested in the proximity of~$\uu=0$, we use the fact that for~$|x|<1$ 
we have~$\overline{B}_2(x)=x^2-|x|+\frac{1}{6}$, to simplify the result to
\be
\frac{\i\pi}{R} \, \Bigl((\alpha \cdot \uu)^2  + \frac{1}{6}\Bigr) \,.
\ee

Upon putting all the pieces together, we obtain the total one-loop correction to the 
Wilsonian action of the three-dimensional theory, 
which we call~$V_\text{eff}(\uu)$ (we justify this name below). We have
\be
\begin{split}
  V_\text{eff}(\uu) 
 \= & \sum_\tf\sum_{\rho_\tf\in\mathcal{R}_\tf} \delta S^\tf_\text{1-loop} \, \\
     \= & \i\pi  \sum_{I, \, \rho_I} \,
    \Bigl(\frac{1}{3R^2} \, \overline{B}_3 \bigl(\rho_\tf\cdot \uu - \tfrac{1}{2} n_0 \, r_I \bigr)
    + \frac{r_I-1}{R} \, \overline{B}_2 \bigl(\rho_\tf\cdot \uu -  \tfrac{1}{2} n_0 \, r_I \bigr) \\
& \qquad\qquad\qquad +  \frac{1}{R}  \sum_{\alpha}  \, \bigl((\alpha \cdot \uu)^2  + \tfrac{1}{6} \bigr) 
    +   \bigl( (r_I-1)^2 \, - \tfrac{1}{6} \bigr) \overline{B}_1 
    \bigl(\rho_\tf\cdot \uu -  \tfrac{1}{2} n_0 \, r_I \bigr) \Bigr) \,.
\end{split}
\label{Veffans3d}
\ee

We now localize the path integral of the light gauge multiplet mode that was excluded from 
the sum~\eqref{eq:sumPrimeVec}, using its Wilsonian effective action, which consists of the 
tree-level action coming from the light~$n=-n_0$ mode in 4d, as well as the one-loop 
action~$\delta S_{\text{1-loop}}$ derived above (in the bosonic sector, which is relevant 
for the localization calculation) from integrating out the heavy modes.
It is useful to keep in mind the different but related problem of calculating the partition function of 
superconformal CS theory coupled to matter on~$\CM_3$ \cite{Kapustin:2009kz}, \cite{Willett:2016adv}. 
In that case the theory localizes onto 
arbitrary constant values of the scalar~$\sigma$ and is supported by the auxiliary scalar~$H$. 
The measure including the one-loop determinant of the localizing action in the non-BPS 
directions is\footnote{Compare with Section~5 of~\cite{Aharony:2013dha}, noting that for 
squashed~$S^3$ with squashing parameter~$b$ one has~$\omega_{1}^{\text{thf}}=\i b$, 
$\omega_{2}^{\text{thf}}=\i b^{-1}$. We leave the derivation of~\eqref{eq:thfModuli} from the 
metric~\eqref{3dmetric1} to future work.} 
\be
\int
\frac{D \underline{\sigma}}{\sqrt{- \omega_1^{\text{thf}}\, \omega_2^{\text{thf}}}}\, 
\prod_{\alpha_+}4\sinh\bigl( \frac{\pi \alpha_+ \cdot \underline{\sigma}}{-\i\, \omega_1^{\text{thf}}} \bigr)
\sinh\bigl( \frac{\pi \alpha_+ \cdot \underline{\sigma}}{-\i\,  \omega_2^{\text{thf}}} \bigr) \,,
\label{1loopM3}
\ee
with~$\omega_{1,2}^{\text{thf}}$ the moduli of the transversely holomorphic foliation 
(THF)~\cite{Closset:2013vra} of~$\mathcal{M}_3$, which we expect to be
\be
 \omega_{1}^{\text{thf}}\=\omega_{2}^{\text{thf}}\=\i\sqrt{\frac{1-\Omega}{1+\Omega}}.\label{eq:thfModuli}
\ee

Recalling from~\eqref{AYval3d} that~$\sigma^i \= u_i/R$, and adding the contribution 
from~$\delta S_{\text{1-loop}}$ in~(\ref{Veffans3d}) (which although arises at one-loop 
in high-temperature EFT, contributes as a ``classical'' piece in the localization computation), 
we obtain the final result for the 
three-dimensional partition function 
\be
Z(\t) \= \int \frac{D\uu}{(-\i\, \tau)^{\mathrm{rk}(G)}}\, 
\prod_{\alpha_+}4\sinh^2\Bigl(\frac{\pi \alpha_+ \cdot \uu}{-\i\, \tau}\Bigr)  \exp \bigl( - V_\text{eff}(\uu) \bigr) \,.
\label{Z3d}
\ee
Noting that the supersymmetric partition function and the Hamiltonian
index are related as \cite{Ardehali:2015hya,Assel:2015nca}
\be
Z(\t) \= e^{2\pi\i\tau E_\text{susy}} \, \CI (\t) \,,
\ee
we see that the result~\eqref{Z3d} agrees precisely with the microscopic
result~\eqref{eq:In0semi-simpleAsy0}--\eqref{Esusy}.

We emphasize that while the above derivation of~$V_{\text{eff}}$ in~\eqref{Veffans3d} 
applies to~$\uu$ near 0, it can be easily extended to generic finite~$\uu$ by modifying the 
vector multiplet discussion. For generic~$\uu$, the non-Cartan components of the~$n=-n_0$ 
mode of the vector multiplet are also heavy, and ought to be integrated out. Consequently the 
sum in~\eqref{eq:sumPrimeVec} would no longer have a prime, and we end up with~$V^r_1$ 
as in~\eqref{eq:V1ren} rather than~$V_1$ in~\eqref{Veffans3d}. This is the EFT derivation of 
the finite-$\uu$ potentials~$V_{1,2}$ found microscopically in~\cite{Cabo-Bizet:2019osg}.

On the other hand, when $n_0=0$, the small-$\uu$ discussion leading up to~\eqref{Veffans3d} 
needs to be modified because now the chiral multiplets have light modes (corresponding to $n=0$). 
As in the discussion around~\eqref{eq:sumPrimeVec} the light mode should be removed from the 
KK sum and instead be included in the dynamical part (to be localized). Indeed, it is well-known that 
for~$n_0=0$ the constant piece of the small-$\tau$ expansion coming from the~$\uu=0$ saddle 
contains the (localized)~$S^3$ partition function of the dimensionally reduced chiral as well as 
vector multiplets~\cite{Ardehali:2015bla} 
(see~\cite{Dolan:2011rp,Spiridonov:2012ww,Niarchos:2012ah,Gadde:2011ia,Imamura:2011uw} 
for earlier work on the connection between 4d indices and $S^3$ partition functions). 

A technical remark is in order regarding our EFT derivation of~\eqref{Z3d}. To reproduce the desired asymptotics, 
we have sent~$\varepsilon(=\frac{\rr}{\tau}=\frac{2\pi\i}{\Omega-1})\to0$, and hence~$\Omega\to\infty$, when 
evaluating the CS actions in Appendix~\ref{sec:CSactionvals}. 
It would be interesting to have a formula of the  type~\eqref{Z3d} for~$\rr\to0$ at finite~$\varepsilon$, which 
would imply that~$\Omega$ and the resulting 3d geometry would be finite-sized.\footnote{The recent 
work~\cite{Cassani:2021fyv} presents such a derivation, although using a background different from ours. 
The precise relation between the two backgrounds is not clear to us at the moment.}

Finally, as  discussed in Appendix~\ref{sec:diffomegas}, we find that the 
effective potential for~$\t$ and~$\s$ not necessarily equal is given by 
making the replacement 
\be
\frac{1}{R^2} \to \frac{1}{\t \, \s} \,, \qquad \frac{1}{R} \to \frac{\t + \s}{2\t \, \s}
\ee
in the effective potential~\eqref{Veffans3d}. The singular pieces are indeed in agreement 
with the microscopic calculations reported 
in~\cite{Kim:2019yrz,Cabo-Bizet:2019osg}.

\subsection{Rational points}

We now turn our attention to the limit of~$\t$ approaching a rational point. 
In the discussion of the previous subsection we used the fact that the radius of the circle~$R$ 
equals~$\t$ which becomes small in the limit, so that we could use an effective three-dimensional description. 
Now we are interested in~$\wt \t = m\t +n \to 0$, with~$n, m \in \IZ$ (with no common factor) 
as in~\cite{Cabo-Bizet:2019eaf}. 
In terms of the variable~$\wt \t$ we have that~$\omega = 2 \pi \i \t = 2 \pi \i (\wt \t -n)/m$ so that
\be
\O \= 1+ \frac{\omega}{\rr} \= 1- \frac{2 \pi \i n}{m\rr} + \frac{2 \pi \i \wt \t}{m\rr} \,,
\ee
and the four-dimensional metric background~\eqref{4dbackgnd} is now
\be \label{4dbackgndrat}
\begin{split}
\dd s_4^2 \= & \dd t_E^2 + \dd \theta^2 
+ \sin^2 \theta \, 
\Bigl(\dd \phi_1 - \frac{2 \pi n}{m\rr} \, \dd t_E -\i \, \bigl(1 + \frac{2 \pi \i \wt \t}{m\rr}\bigr) \, 
\dd t_E \Bigr)^2 \\
& \qquad \qquad \qquad + \cos^2 \theta \, 
\Bigl(\dd \phi_2 - \frac{2 \pi n}{m\rr} \, \dd t_E -\i \, \bigl(1+ \frac{2 \pi \i \wt \t}{m\rr}\bigr) \, 
\dd t_E \Bigr)^2  \,.
\end{split}
\ee
In terms of the following new coordinates and new parameters,
\be
\wt \rr \= m \rr \,, \qquad 
\wt \O \= 1+ \frac{2 \pi \i \wt \t}{\wt \rr}  \,,  \qquad 
\wt \phi_i \= \phi_i - \dfrac{2 \pi n}{\wt \rr} t_E \,,
\ee
the above metric is  
\be \label{4dbackgndrat1}
\dd s_4^2 \=  \dd t_E^{2} + \dd \theta^2 
+ \sin^2 \theta \, 
\bigl(\dd \wt \phi_1  -\i \, \wt \O  \, \dd t_E \bigr)^2 + \cos^2 \theta\,  \bigl(\dd \wt \phi_2  -\i \, \wt \O  \,\dd t_E \bigr)^2  \,,
\ee
with~$\wt \phi_1$, $\wt \phi_2$ being $2\pi$-periodic as before, 
and the periodic identification going around the time circle is 
\be \label{newident}
\bigl(t_E \,, \;  \wt \phi_1 \,, \;  \wt \phi_2 \bigr) \; \sim \;
\Bigl( t_E + \frac{\wt \rr}{m} \,, \;  \wt \phi_1 - \frac{2 \pi n}{m} \,, \;  \wt \phi_2 - \frac{2 \pi n}{m} \Bigr) \,.
\ee
The metric configuration~\eqref{4dbackgndrat1} with the identifications~\eqref{newident} is 
simply a global identification, or orbifold, of the configuration considered in the previous subsection with 
the new parameters~$(\wt \rr, \wt\t, \wt \O)$ replacing $(\rr,\t, \O)$.~\footnote{We learned about 
these orbifolds from a talk by O.~Aharony at the Stony Brook seminar series in November 
2020~\cite{AhaSBTalk}.}

On the covering space, 
going around the time circle shifts~$\wt t_E \to \wt t_E + \wt \rr$ and~$\wt \phi_i \to \wt \phi_i + 2 \pi n$.
The latter identification can be trivialized by using the independent~$2\pi$-periodicity of~$\wt \phi_i$, so that 
we have the identification~$\bigl(\,\wt t_E \,,  \wt \phi_1 \,,  \wt \phi_2 \, \bigr) \sim 
\bigl(\, \wt t_E + \wt \rr \,,   \wt \phi_1  \,,   \wt \phi_2 \, \bigr)$. 
On this configuration we can perform the dimensional reduction to three dimensions.
The relevant considerations of the previous subsection go through exactly as before 
with the replacement~$(\rr,\t, \O) \mapsto (\wt \rr, \wt\t, \wt \O)$. Actually, because the gauge 
holonomies on the cover wrap a circle~$m$ times larger than the original~$S^1$, we also get 
a replacement~$u_j\to mu_j$. Moreover, since~$\xi_I$ (which equals~$-n_0r_I/2$ for~$(m,n)=(1,0)$) 
effectively plays the role of a flavor chemical potential in our problem as mentioned around~\eqref{eq:defXiGen}, 
we expect a similar replacement~$-n_0r_I/2\to m\xi_I$. We can see this replacement arise more directly as follows.

We multiply the first term in~\eqref{defSf} by~$\frac{m^2}{m^2}$, and the second term by~$\frac{m}{m}$. 
This amounts to~$\rr\to m\rr$ and~$u_j\to mu_j$ as mentioned above, but also~$k_Y\to mk_Y$ (which 
corresponds to keeping only the singlet modes under the $\mathbb{Z}_m$ quotient) as well as~$n_0\to m n_0$. 
On the other hand, writing~$A^{\text{nm}}_Y$ in~\eqref{AYVYval} in terms of~$\wt{\tau}$ instead of~$\tau$ 
amounts to yet another replacement~$n_0\to n_0+\frac{2n}{m}$. Combining these two effects yields the 
desired~$-n_0r_I/2\to m\xi_I$ replacement.

With the preceding substitutions in the results of the previous 
subsection, we thus arrive at the potentials~$\wt V_{2,1}$ in~\eqref{eq:defVsRat}. 
We then take the~$\IZ_m$ quotient which has two effects as usual. 
Firstly it reduces the volume of the three-dimensional space, and 
secondly it introduces new topologically non-trivial sectors in the path integral over the 
gauge-field configurations.
The change in calculations involving local gauge-invariant Lagrangians will therefore 
be only a reduction in the action by a factor of~$m$. 
This explains the reduction of the effective potential by a factor of~$m$ as in~\eqref{eq:In0semi-simpleAsyRat0}.

Finally we discuss the constant terms (in~$\wt \t$) arising from the functional integral over the 
dynamical gauge multiplet. 
There are a few subtleties. Firstly the actions like the gravitational CS action will depend on the 
global properties of the orbifold. 
Then we need to calculate the partition function of the orbifold space with a background graviphoton.
Assuming as in the previous subsection that the expected THF moduli arise, and that by 
re-scaling and contour deformation (as discussed around~\eqref{eq:Z0Rational}) the THF
moduli can be replaced with those of round~$S^3$, the calculation presumably reduces to 
an~$S^3/\mathbb{Z}_m$ partition function as in~\cite{Benini:2011nc,Alday:2012au,Gang:2019juz,Willett:2016adv}, 
with the~$\IZ_m$ action following from~\eqref{newident} to be
\be
(\wt\phi_1,\wt\phi_2) \; \sim \; (\wt\phi_1-\frac{2\pi n}{m},\wt\phi_2-\frac{2\pi n}{m}),
\ee
which for~$n=1$ coincides with that of the lens space~$L(m,-1)$.
Here one has to be careful about how the measure on the space 
of constant scalars~$\s_i$ is affected by the  four-dimensional orbifold~\eqref{newident}. 
We leave these interesting questions to future work, noting that 
the result of these considerations indeed agrees with the microscopic 
answer~\eqref{eq:N=4indexAsyRational}, with the~$\rO(\wt \t)$ piece explained by the 
supersymmetric Casimir energy factor as before.

\vspace{0.4cm}

\noindent \textbf{Note Added.} The paper~\cite{Cassani:2021fyv}, which appeared on the arXiv 
the same day as the first version of this paper, has some overlap with our section~\ref{sec:4dto3d}. 
The paper~\cite{Jejjala:2021hlt}, which appeared on the arXiv soon after, has some overlap 
with our section~\ref{sec:SCI}.

\section*{Acknowledgements}

We would like to thank Alejandro Cabo-Bizet, Daniel Butter, Davide Cassani, Cyril Closset, Zohar Komargodski, Neil Lambert, Stavros Garoufalidis,
Bernard de Wit, and Don Zagier for useful discussions and comments.
This work is supported by the ERC Consolidator Grant N.~681908, ``Quantum black holes: A macroscopic 
window into the microstructure of gravity'', and by the STFC grant ST/P000258/1. 
AAA would like to especially thank Junho~Hong for several helpful discussions and 
collaboration on a related project.

\appendix

\section{Asymptotic estimates of the special functions \label{app:Estimates}}

\subsection{$\tau\to0$}

We first consider the limit $\tau\to0$. More precisely, in the rest of this subsection 
we assume~$\mathrm{arg}(\t)$ is in compact domains avoiding integer multiples 
of~$\frac{\pi}{2}$ as~$|\tau|\to0$.

For the Pochhammer symbol $(q;q)$ the small-$\tau$ asymptotics is standard:
\begin{equation}
    (q;q) \; \simeq \;  
    \frac{1}{\sqrt{-\i\,\tau}} \;  \exp \Bigl( -\frac{2\pi \i}{24\, \tau}  -\frac{2\pi \i \,\tau}{24} \Bigr) 
    \qquad(\text{as }|\tau|\to0) \,.
\label{eq:PochEst}
\end{equation}
Recall that the symbol $\simeq$ means that logarithms (on appropriate branches) 
of the two sides (assumed to be non-zero) 
are equal to all orders in the small parameter (here in $|\tau|$). 

For the chiral multiplet elliptic gamma functions we have the following estimate, 
valid for any $r\in\mathbb{R}$, uniformly in $z$ over compact subsets 
of~$\mathbb{R}\setminus\mathbb{Z}$ (see Proposition~2.11 of~\cite{Rains:2006dfy} 
or Equation~(3.53) of~\cite{Ardehali:2015bla}):
\begin{equation}
\begin{split}
    \Ge(r\tau+z)& \;\simeq \; \exp \biggl(-2\pi \i \, \biggl(\,\frac{\overline{B}_3(z)}{6\tau^2}
    +(r -1) \, \frac{\overline{B}_2(z)}{2\tau}+\frac{(r-1)^2-\frac{1}{6}}{2} \, \overline{B}_1(z) 
    +\frac{(r-1)^3-\frac{r-1}{2}}{6}\,\tau \,\biggr) \biggr) \,,
\label{eq:numEst}
\end{split}
\end{equation}
as $|\tau|\to0$.
Here $\overline{B}_j(z)$ are the \emph{periodic Bernoulli polynomials} defined, for~$z \in \IR$
through their Fourier series expansion, 
\begin{equation}
   -\frac{(2\pi \i)^j}{j!} \, \overline{B}_j(z) \=  \sum_{k\in\mathbb{Z}}{}^{'}\;\frac{\rme^{2\pi \i k z}}{k^j}
    \qquad (z\in\mathbb{R}\,,\ j\ge1) \,.
\label{eq:FourierBernoulli}
\end{equation}
The prime in the above formula means that $k=0$ has to be omitted, and that in the $j=1$ 
case---where the series is not absolutely convergent---the sum is in the sense of Cauchy principal value.

For~$x \in \mathbb{R}\setminus\mathbb{Z}$, we have~$\overline{B}_j(x)=B_j(\{x\})$ 
with~$\{\cdot\}:=\cdot-\lfloor\cdot\rfloor$ the fractional-part 
function. When $j>1$ this also holds for $x\in\mathbb{Z}$ 
(and so $\overline{B}_j(\mathbb{Z})=B_j(0)$). When $j=1$ on the other hand 
$\overline{B}_1(\mathbb{Z})=0$, while $B_1(0)=-1/2$.

The Bernoulli polynomials are uniquely characterized by
\begin{equation}
    B_0(u)\=1,\qquad B'_j(u)\=jB_{j-1}(u),\qquad B_j(0)\=B_j(1)\quad \text{for $j>1$}\,,
    \label{eq:recursiveBernoulli}
\end{equation}
and the first three non-trivial ones are explicitly
\begin{equation}
\begin{split}
    B_1(x)&\=x-\frac{1}{2} \,,\\
    B_2(x)&\=x^2-x+\frac{1}{6} \,,\\
    B_3(x)&\=x^3-\frac{3}{2}x^2+\frac{1}{2}x \,.
\end{split}
\end{equation}

The connection between $\overline{B}_j$ and the Bernoulli polynomials can be verified 
by first noting that for $j=1$ the left-hand side of \eqref{eq:FourierBernoulli} is 
essentially the Taylor expansion of the logarithm function, and then observing that $B_j$ 
are uniquely characterized by
\begin{equation} \nn
    B_0(u)\=1,\qquad B'_j(u)\=jB_{j-1}(u),\qquad B_j(0)\=B_j(1)\quad \text{for $j>1$}\,.
\end{equation}

With the aid of (\ref{eq:FourierBernoulli}) one can easily prove relations such as
\begin{equation}
   \sum_{\ell=1}^{C-1}\overline{B}_3 \Bigl(x+\frac{\ell}{C} \Bigr)\=
   \frac{\overline{B}_3(Cx)}{C^{2}}-\overline{B}_3(x)\,,\qquad(\text{Raabe's formula})\label{eq:Raabe} 
\end{equation}
and
\begin{equation}
\begin{split}
    \sum_{\ell=1}^{n-1}\Bigl(\overline{B}_{2} \Bigl(x+m\frac{\ell}{n} \Bigr)
    -\overline{B}_{2} \Bigl(x-m\frac{\ell}{n} \Bigl) \Bigr)&\=0\,,\\
    \sum_{\ell=1}^{n-1}\Bigl(\overline{B}_2 \Bigl(\frac{\ell}{n} \Bigr)\,\overline{B}_{2} \Bigl(x+m\frac{\ell}{n} \Bigr)
    -\overline{B}_2\Bigl(\frac{\ell}{n}\Bigl) \, \overline{B}_{2}\Bigl(x-m\frac{\ell}{n}\Bigr)\Bigl)&\=0 \,,
\label{eq:BernoulliBarIdentities}
\end{split}
\end{equation}
valid for $m,n\in\mathbb{Z}_{>0}$ relatively prime and $x\in\mathbb{R}$, by using the Fourier 
expansion of the Bernoulli functions, and swapping the sum over Fourier modes with the 
sum over~$\ell$.\footnote{Note that similar 
operations with $\overline{B}_1$ are not allowed, because its Fourier expansion is not absolutely 
convergent. This is the source of sophistication of the Dedekind sum defined below in terms 
of~$\overline{B}_1$---or more specifically the source of the nontrivial dependence 
of~(\ref{eq:ourDedekind Sum}) on $n$. 
(Readers familiar with Eisenstein series might recall
similar ``anomalous'' behavior from $E_2$ and its associated elliptic functions.) 
A closely 
related fact is that $\overline{B}_{j>1}$ are continuous, but $\overline{B}_1$ has discontinuities 
on $\mathbb{Z}$.}

The estimate (\ref{eq:numEst}) is particularly useful for the chiral multiplet gamma functions 
in~(\ref{defVmicro}) when the integral is dominated by the 1-center holonomy configurations 
with~$z_i-z_j=0$. This is because the complex phase~$2\pi n_0/3$ shifts the argument of the 
chiral multiplet gamma functions safely into the interior of the domain~$z\in\mathbb{R}\setminus\mathbb{Z}$ 
where the estimate is uniformly valid. On the other hand, since the vector multiplet gamma functions 
in~(\ref{defVmicro}) lack such phase shifts in their arguments, the estimate~(\ref{eq:numEst}) is not 
appropriate for them near the 1-center holonomy configurations when~$\tau\to0$.

The estimate \eqref{eq:numEst} is not uniformly valid, with respect to $z$, over intervals 
containing $\mathbb{Z}$. There is a well-known improvement of it around $z=0$ however, 
which is valid uniformly over compact subsets of $(-1,1)$, and we will use for vector 
multiplet elliptic gamma functions in the index. It reads (see Proposition~2.10 
of~\cite{Rains:2006dfy} or Equation~(2.16) of~\cite{Ardehali:2015bla})
\begin{equation}
\begin{split}
\Gamma_\rme(r\tau+z)\simeq \rme^{2\pi i
R_{0}(r\tau+z;\tau)}\, \Gamma_h(r\tau+z;\tau,\tau),\label{eq:GammaToHypGamma}
\end{split}
\end{equation}
where
\begin{equation}
    R_0(z;\tau)=-\frac{z^3}{6\tau^2}+\frac{z^2}{2\tau}-\frac{(1+5\tau^2)z}{12\tau^2} + \frac{1}{12\tau} + \frac{\tau }{12},
\end{equation}
and $\Gamma_h(x;\omega_1,\omega_2)$ is the hyperbolic gamma function.

Using the estimate \eqref{eq:GammaToHypGamma} and the ``product formula''
\begin{equation}
    \frac{1}{\Gamma_h(x;\omega_1,\omega_2)\Gamma_h(-x;\omega_1,\omega_2)}
    \=-4\sin \big(\frac{\pi x}{\omega_1}\big)\sin \big(\frac{\pi x}{\omega_2}\big),\label{eq:gammaHprodGen}
\end{equation}
the next estimate follows
(cf.~Equation~(2.18) of~\cite{Ardehali:2015bla}):
\begin{equation}
    \frac{1}{\Ge(z)\, \Ge(-z)} \;\simeq \; \rme^{-4\pi \i \,R^+_0(z;\tau)} \;
    4\sin \Bigl(\frac{\pi z}{\tau}\Bigr) \, \sin \Bigl(-\frac{\pi z}{\tau}\Bigr),
\label{eq:denomEstSins}
\end{equation}
valid uniformly in $z$ over compact subsets of~$(-1,1)$, with
\begin{equation}
    R^+_0(z;\tau) \defeq \frac{R_0(z;\tau)+R_0(-z;\tau)}{2}= \frac{z^2}{2\tau} + \frac{1}{12\tau} + \frac{\tau }{12} \,.
\label{eq:R0}
\end{equation}

Note that since $\Ge(z+2\tau)\, \Ge(-z+2\tau)=\frac{1}{\Ge(z)\, \Ge(-z)}$, 
and $\sin(\i \, x)=\i\sinh(x)$, we can write \eqref{eq:denomEst} alternatively as
\be
    \Ge(z+2\tau)\, \Ge(-z+2\tau)\;\simeq \; \rme^{-4\pi \i \,R^+_0(z;\tau)} \, 
    4 \sinh^2 \Bigl(\frac{\pi z}{-\i\tau}\Bigr) \,.\label{eq:denomEst}
\ee

While we have presented two separate estimates~(\ref{eq:numEst}) and~(\ref{eq:denomEst}) 
for the chiral and vector multiplet gamma functions, both of them can in fact be derived from the 
``central estimate'' \eqref{eq:GammaToHypGamma}. 
Deriving~(\ref{eq:numEst}) from the central estimate requires only an extra step to simplify the 
hyperbolic gamma functions arising from \eqref{eq:GammaToHypGamma} using Corollary~2.3 of~\cite{Rains:2006dfy}, 
as explained in Proposition~2.11 there.

\subsection{$\tau\to\mathbb{Q}$}

We now consider
\begin{equation}
\wt \tau \;\equiv \; m\tau+n \to 0 \,, 
\end{equation}
with $m,n$ relatively prime. More precisely, in the rest of this subsection we 
assume that~$\mathrm{arg}(\wt{\tau})$ is in compact domains avoiding integer 
multiples of~$\pi/2$ as~$|\wt{\tau}|\to0$.

To obtain the asymptotics of the Pochhammer symbol we note that for integer $a,b,c,d$ 
satisfying $ad-bc=1$ with $c>0$, we have
\begin{equation}
    \eta\left(\frac{a\tau+b}{c\tau+d}\right) \=
    \exp \Bigl(2\pi \i \Bigl(\frac{a+d}{24c}-\frac{1}{8}-\frac{s(d,c)}{2} \Bigr)\Bigr) \, 
    (c\tau+d)^{1/2} \, \eta(\tau) \,,
\label{eq:modularEta}
\end{equation}
with $s(d,c)$ the Dedekind sum
\begin{equation}
    s(d,c)\=\sum_{\ell=1}^{c-1} \, \frac{\ell}{c} \, \overline{B}_1\big(d\frac{\ell}{c}\big) \,.
\label{eq:DedSumDef}
\end{equation}
Since the gcd$(m,n) =1$, there exist integers~$a$, $b$ such that~$an-bm=1$. 
Now we use~(\ref{eq:modularEta}) with~$(c,d)=(m,n)$. Noting that
$a \t + b = a(\wt \t -n)/m + b = a \wt \t/m -1/m$, we obtain
\begin{equation}
    (q;q)\;\simeq \; \frac{1}{\sqrt{-\i \wt{\tau}}}  \;
      \exp \Bigl( -\frac{2\pi \i}{24m\wt{\tau}} - \frac{2\pi \i \, \wt\tau}{24m} + \i\pi s(n,m) \Bigr) \,,
\label{eq:PochRationalEst}
\end{equation}
in the limit of our interest.
Our Dedekind sum is explicitly
\begin{equation}
    s(n,m)\=\sum_{\ell=1}^{m-1}\frac{\ell}{m} \,\overline{B}_1\big(n\frac{\ell}{m}\big) \,.
\label{eq:ourDedekind Sum}
\end{equation}

To obtain an estimate for the elliptic gamma function we first note the identity \cite{felder2000elliptic}
\begin{equation}
    \Gamma(\zeta;q,q)\=\prod_{\ell=0}^{2(m-1)}\Gamma(\zeta q^\ell;q^m,q^m)^{m-|\ell-(m-1)|}
    \=\prod_{\ell=0}^{2(m-1)}\Gamma(\zeta \rme^{-2\pi \i n\frac{\ell}{m}}
     \wt{q}^{\frac{\ell}{m}};\wt{q},\wt{q})^{m-|\ell-(m-1)|},\label{eq:gammaFVprod}
\end{equation}
with $\wt{q}=\rme^{2\pi \i \wt{\tau}}$.

Using (\ref{eq:numEst}) on the right-hand side of (\ref{eq:gammaFVprod}) we get
\begin{equation}
\begin{split}
   & \frac{1}{2\pi \i \,}\log\Ge(z) \; \sim \\
   &-\frac{1}{6\wt{\tau}^2} \biggl(\,\sum_{\ell=1}^{m-1}\ell
   \Bigl(\overline{B}_3 \Bigl(z+\frac{n}{m}-n\frac{\ell}{m} \Bigr)
   +\overline{B}_3 \Bigl(z+\frac{n}{m}+n\frac{\ell}{m} \Bigr)\Bigr)
   +m\,\overline{B}_3\Bigl(z+\frac{n}{m} \Bigr) \biggr)\\
    &\ -\frac{1}{2\wt{\tau}} \biggl(\, \sum_{\ell=1}^{n-1}\ell\Bigl( \Bigl(\frac{\ell-1}{m}-1 \Bigr) \,
    \overline{B}_2 \Bigl(z+\frac{n}{m}-n\frac{\ell}{m} \Bigr)+ \Bigl(\frac{2m-\ell-1}{m}-1 \Bigr) \,
    \overline{B}_2 \Bigl(z+\frac{n}{m}+n\frac{\ell}{m} \Bigr) \Bigr)\\
    &\ \ \  +m \,\Bigl(\frac{m-1}{m}-1 \Bigr) \,\overline{B}_2\Bigl(z+\frac{n}{m}\Bigr) \biggr)\\
    &\ -\frac{1}{2} \biggl( \, \sum_{\ell=1}^{m-1}\ell  \Bigl( \Bigl(\frac{\ell-1}{m}-1 \Bigr)^2
    -\frac{1}{6} \Bigr) \, \overline{B}_1 \Bigl(z+\frac{n}{m}-n\frac{\ell}{m} \Bigr)+ \Bigl( \Bigr(\frac{2m-\ell-1}{m}-1\Bigr)^2
    -\frac{1}{6} \Bigr)\,\overline{B}_1 \Bigl(z+\frac{n}{m}+n\frac{\ell}{m} \Bigr) \Bigr)\\
    &\ \ \ +m \,   \Bigl( \Bigl(\frac{m-1}{m}-1 \Bigr)^2-\frac{1}{6}\Bigr) \,\overline{B}_1 \Bigl(z+\frac{n}{m} \Bigr) \biggr)\\
    &\ - \wt{\tau} \, \biggl( \, \sum_{\ell=1}^{m-1}\ell \, \Bigl(\frac{1}{6}\,  \Bigl(\frac{\ell-1}{m}-1 \Bigr)^3
    -\frac{1}{12}\, \Bigl(\frac{\ell-1}{m}-1 \Bigr)
    +\frac{1}{6} \, \Bigl(\frac{2m-\ell-1}{m}-1 \Bigr)^3 -\frac{1}{12} \, \Bigl(\frac{2m-\ell-1}{m}-1 \Bigr) \Bigr) \\
    &\ \ \ +m \, \Bigl(\frac{1}{6} \, \Bigl(\frac{m-1}{m}-1 \Bigr)^3 -\frac{1}{12}\, \Bigl(\frac{m-1}{m}-1 \Bigr) \Bigr)\biggr) \,.
    \end{split}\label{eq:longRationalEst}
\end{equation}
Now using the identity\footnote{See Equation~(4.54) in \cite{Cabo-Bizet:2019eaf} or Equation~(3.12) 
in~\cite{ArabiArdehali:2019orz}. A simple proof is possible via (\ref{eq:FourierBernoulli}).}, 
for gcd($m,n$)$=1$, $k>1$, 
\begin{equation}
    \sum_{\ell=1}^{m-1}\ell \, \Bigl( \overline{B}_k \, \Bigl(x-n\frac{\ell}{m} \Bigr)
    +\overline{B}_k \, \Bigl(x+n\frac{\ell}{m} \Bigr)\Bigr)+m \, \overline{B}_k(x)
    \= \frac{1}{m^{k-2}} \, \overline{B}_k(mx) \,,
\label{eq:remarkableId}
\end{equation}
and (\ref{eq:FourierBernoulli}), we can simplify (\ref{eq:longRationalEst}) to
\begin{equation}
   \Ge( z) \; \simeq \; \exp \biggl(-\frac{2\pi \i}{m} \Bigl(\, \frac{\overline{B}_3(mz)}{6 \wt{\tau}^2}
   -\frac{\overline{B}_2(mz)}{2 \wt{\tau}}+C(m,n,z) -\frac{1}{12} \wt{\tau} \Bigr) \biggr),
\label{eq:gammaRationalEst}
\end{equation}
for $mz\in\mathbb{R}\setminus\mathbb{Z}$, as $\wt{\tau}\to0$. 
Here $C(m,n,z)$ stands for ($-m$ times) the fourth and fifth lines 
of the right-hand side of~(\ref{eq:longRationalEst}).

Generalizing the above derivation in a straightforward manner leads to
\begin{equation}
\begin{split}
  & \Ge( z+r\tau)\simeq \\
  & \quad \exp \biggl(-\frac{2\pi \i}{m} \Bigl(\frac{\overline{B}_3(mz-nr)}{6 \wt{\tau}^2}
  +(r-1)\, \frac{\overline{B}_2(mz-nr)}{2 \wt{\tau}}+C(m,n,z,r) 
  +\frac{(r-1)^3-\frac{r-1}{2}}{6} \wt{\tau} \Bigr) \biggr) \,,
\label{eq:gammaRationalEstWithR}
\end{split}
\end{equation}
for $r\in\mathbb{R}.$ This is the analog of (\ref{eq:numEst}) for $\wt\tau\to0$.

The explicit expression for $C(m,n,z,r)$ is
\begin{equation}
\begin{split}
    C(m,n,z,r)&\=-\frac{m}{2}\big[\sum_{\ell=1}^{m-1}\ell\big(((\frac{\ell+r-1}{m}-1)^2
    -\frac{1}{6})\overline{B}_1(z+\frac{n}{m}-n\frac{\ell+r}{m})\\
    &\ \ \ +((\frac{2m-\ell+r-1}{m}-1)^2-\frac{1}{6})\overline{B}_1(z+\frac{n}{m}+n\frac{\ell-r}{m})\big)\\
    &\ \ \ +m((\frac{m+r-1}{m}-1)^2-\frac{1}{6})\overline{B}_1(z+\frac{n}{m}-\frac{nr}{m})\big] \,.
\label{eq:C(m,n,r,z)}
\end{split}
\end{equation}

The estimate~(\ref{eq:gammaRationalEstWithR}) is important to derive our results for 
the asymptotic expansion of the index near the roots of unity. 
It is valid uniformly over compact subsets 
of $z\in\mathbb{R}\setminus\frac{\mathbb{Z}}{m}$, because using (\ref{eq:numEst}) 
on the right-hand side of (\ref{eq:gammaFVprod}) is allowed only 
if $z-n\frac{l}{m}\notin\mathbb{Z}$ for $\ell=0,\dots,m-1$. 

We can also use (\ref{eq:gammaFVprod}) for $z$ near $0$. More precisely,  on the RHS of 
(\ref{eq:gammaFVprod}), for fixed~$z \in (-\frac{1}{m}, \frac{1}{m})$,
we can use \eqref{eq:denomEst} for the $\ell=0,m$ terms, and use \eqref{eq:numEst} for all other $\ell$. 
With the aid of the ``reflection formula''
\begin{equation}
  \Gamma_h(x+\frac{\omega_1+\omega_2}{2};\omega_1,\omega_2)
  \Gamma_h(-x+\frac{\omega_1+\omega_2}{2};\omega_1,\omega_2)=1.\label{eq:gammaHrefl}
\end{equation}
which gets rid of the hyperbolic gammas arising from $\ell=m$, and using the 
``product formula'' \eqref{eq:gammaHprodGen} to trade the hyperbolic gammas 
arising from $\ell=0$ for hyperbolic sines, we obtain
\begin{equation}
    \frac{1}{\Ge(z)\Ge(-z)}\;\simeq \;  \exp \Bigl( -\frac{4\pi \i}{m} \wt{R}^+_0(z; \wt{\tau}) +4\pi\i s(n,m)\Bigr) \,
    4\sinh^2 \Bigl(\frac{\pi z}{-\i\wt\tau}\Bigr) \,,
\label{eq:denomEstRational}
\end{equation}
where
\begin{equation}
    \wt{R}^+_0(z; \wt{\tau}) \defeq \frac{m^2 z^2}{2 \wt{\tau}} 
    + \frac{1}{12 \wt{\tau}} +  \frac{\wt{\tau}}{12} \,.
\label{eq:R0Rational}
\end{equation}
This is the analog of (\ref{eq:denomEst})--(\ref{eq:R0}) for $\wt\tau\to0$, 
and is similarly useful (i.e.~uniformly valid) in a neighborhood of $z=0$.

An estimate similar to \eqref{eq:denomEstRational} for $z$ near general nonzero $\frac{\mathbb{Z}}{m}$
can be obtained as well. We focus for simplicity on the $n=1$ case (i.e. $\tau\to-\frac{1}{m}$). 
We write $z=\ell_0/m+z'$ and appeal to \eqref{eq:gammaFVprod}. We have to use the 
estimate \eqref{eq:GammaToHypGamma} for $\ell=\ell_0,\ell_0+m$, and the 
estimate \eqref{eq:numEst} for all other $\ell$ in the product \eqref{eq:gammaFVprod} 
for $\Gamma_\rme(z)$. Similarly we have to use the estimate \eqref{eq:GammaToHypGamma} 
for $\ell=-\ell_0+m,-\ell_0+2m$, and the estimate \eqref{eq:numEst} for all other $\ell$ in the 
product for $\Gamma_\rme(-z)$. The result is (up to a constant phase that we suppress)
\begin{equation}
\begin{split}
  \frac{1}{\Ge(z)\Ge(-z)}\;\simeq \; 
  & \exp\Bigl(-2\pi \i [\frac{\frac{1}{6}+m^2z'^2}{m\wt\t}-\frac{1}{2}+\frac{\wt\t}{6m}] \Bigr) \; \times \\
  & \qquad     \bigl[\Gamma_h(z'+\frac{\ell_0}{m}\wt{\tau} ,\wt{\tau},\wt{\tau})^{\ell_0+1} 
  \Gamma_h(-z'+\frac{m-\ell_0}{m}\wt{\tau} ,\wt{\tau},\wt{\tau})^{m-\ell_0+1}\\
  &\qquad \Gamma_h(z'+\frac{\ell_0+m}{m}\wt{\tau} ,\wt{\tau},\wt{\tau})^{m-\ell_0-1}
  \Gamma_h(-z'+\frac{2m-\ell_0}{m}\wt{\tau} ,\wt{\tau},\wt{\tau})^{\ell_0-1}\bigr]^{-1} \,.
\label{eq:denomEstRational0}
\end{split}
\end{equation}
Using the reflection formula \eqref{eq:gammaHrefl} we can simplify the above product of the 
hyperbolic gamma functions to find (up to the neglected constant phase)
\begin{equation}
\begin{split}
  \frac{1}{\Ge(z)\Ge(-z)}\;\simeq \; 
  & \exp\Bigl(-2\pi \i [\frac{\frac{1}{6}+m^2z'^2}{m\wt\t}-\frac{1}{2}+\frac{\wt\t}{6m}] \Bigr) \; \times \\
  & \qquad     \bigl[\Gamma_h(z'+\frac{\ell_0}{m}\wt{\tau} ,\wt{\tau},\wt{\tau})\, 
  \Gamma_h(-z'+\frac{m-\ell_0}{m}\wt{\tau} ,\wt{\tau},\wt{\tau})\bigr]^{-2} \,.
\label{eq:denomEstRational0.5}
\end{split}
\end{equation}
Now we use\footnote{This relation can be proven using the reflection formula \eqref{eq:gammaHrefl} 
together with $\Gamma_h(x+\tau,\tau,\tau)=2\sin\big(\frac{\pi x}{\tau}\big)\Gamma_h(x,\tau,\tau)$.}
\begin{equation}
  \bigl[\Gamma_h(x ,\wt{\tau},\wt{\tau})\, \Gamma_h(-x+\wt{\tau} ,\wt{\tau},\wt{\tau})\bigr]^{-2} 
  \=- 4\sinh^2 \Bigl(\frac{\pi x}{-\i\,\wt\tau}\Bigr),
\end{equation}
to simplify \eqref{eq:denomEstRational0.5} to (up to the neglected constant phase)
\begin{equation}
\begin{split}
  \frac{1}{\Ge(z)\Ge(-z)}\;\simeq \; 
  & \exp\Bigl(-2\pi \i [\frac{\frac{1}{6}+m^2z'^2}{m\wt\t}+\frac{\wt\t}{6m}] \Bigr) \; 
  4\sinh^2 \Bigl(\frac{\pi (z'+\frac{\ell_0}{m}\wt\t)}{-\i\,\wt\tau}\Bigr).
\label{eq:denomEstRational0.75}
\end{split}
\end{equation}

\section{Supersymmetric three-dimensional Chern-Simons actions \label{sec:CSactions}}

In this appendix we present the bosonic part of supersymmetrized three-dimensional 
Chern-Simons actions. 
We work in the context of three-dimensional~$\CN=2$ supersymmetric gauge theory
coupled to off-shell three-dimensional supergravity.  
We first collect all allowed Chern-Simons terms including background and dynamical gauge fields, 
and then write the corresponding supersymmetrizations, 
following the presentation of Appendix~A of~\cite{DiPietro:2016ond}. 
We then evaluate the actions for the field configurations that we 
consider in Section~\ref{sec:4dto3d}.

The CS terms have the form
\be
\frac{1}{\pi^2} \, \int_{\CM_3}\CA_x\wedge \mathrm{d}\CA_y\,,
\ee 
where~$x$ and~$y$ run over all possible gauge fields and with a coupling that we discuss below.
Below we present the bosonic parts of the supersymmetric completions of the various cases $x$-$y$, 
following~\cite{DiPietro:2016ond,Closset:2018ghr}. 
Firstly we have the gauge-gauge and gauge-R CS terms,
\be
\begin{split}
s_{\text{g-g}}^{ij} & \= \frac{1}{\pi^2} \, \int_{\CM_3}\dd^3 x \, \sqrt{g} \,  \Bigl( \epsilon^{\mu\nu\rho} \, 
\CA^i_{\mu} \, \partial_\nu  \CA^j_{\rho}  + 2\i \, \mathcal{D}^i \, \sigma^j  \Bigr) \,,\\
s_{\text{g-R}}^{i} & \= \frac{1}{\pi^2} \, \int_{\CM_3}\dd^3 x \, \sqrt{g}  \, 
\Bigl(\epsilon^{\mu\nu\rho} \CA^i_{\mu} \, \partial_\nu  \, \bigl(\CA^{(R)}_\rho - \tfrac 12 v_\rho \bigr)  +\i  \CD^i H 
+\i \, \frac{ \sigma^i}{4} \bigl(R^{(3)}  + 2 v_\mu v^\mu + 2 H^2 \bigr)  \Bigr)  \,.
\label{eq:CSactions1}
\end{split}
\ee
Here~$i,j$ are Cartan labels for the gauge group, $\CA$ is the three-dimensional gauge field, 
$\CD$ is the~$D$-term auxiliary scalar of the three-dimensional~$\mathcal{N}=2$ vector 
multiplet, $\sigma$ is the (Coulomb branch) scalar of the three-dimensional $\mathcal{N}=2$ 
vector multiplet, $\CA^{(R)}$ is the three-dimensional background gauge field for the $R$ current, 
$v^\mu =-\i \, \ve^{\mu\nu\rho} \, \partial_\nu \, c_\rho$  is the Hodge dual of the 
graviphoton $c_\mu$ (the background $U(1)_{\text{KK}}$ gauge field), $H$ is the scalar in the 
supergravity multiplet, and~$R^{(3)}$ is the Ricci scalar of $\CM_3$. 

Then we have the background R-R CS terms and the gravitational 
CS term for the spin connection~$\omega$,
\be
\begin{split}
R_{\CM_3} & \= \frac{1}{\pi^2} \, \int_{\CM_3} \dd^3x \, \sqrt{g}\, \Bigl(\epsilon^{\mu\nu\rho} \, 
\bigl(\mathcal A^{(R)}_\mu-\tfrac12 v_\mu \bigr) \partial_\nu \bigl(\mathcal A^{(R)}_\rho-\tfrac12 v_\rho \bigr)
+\i\, \frac{H}{2}\left(R^{(3)}+2v_\mu v^\mu+2H^2\right) \Bigr) \,, \\
G_{\CM_3}  & \= \frac{1}{\pi^2} \, \int_{\CM_3} \dd^3x \, \sqrt{g}\, \Bigl( \epsilon^{\mu\nu\rho}
\Tr \bigl(\omega_\mu \, \partial_\nu \, \omega_\rho-\tfrac23 \omega_\mu \, \omega_\nu \, \omega_\rho \bigr)
	+4 \, \epsilon^{\mu\nu\rho} \bigl(\mathcal A^{{(R)}}_\mu-\tfrac32 v_\mu \bigr)
	\partial_\nu \bigl(\mathcal A^{(R)}_\rho-\tfrac32 v_\rho \bigr) \Bigr) \,.
\label{eq:CSactions2}
\end{split}
\ee

Finally we have the CS actions involving the graviphoton. These are gauge-KK, 
R-KK, and KK-KK, whose bosonic parts read \cite{DiPietro:2016ond}
\be
\begin{split}
s_{\text{g-KK}}^{i} & \= \frac{1}{\pi^2} \, \int_{\CM_3}\dd^3 x \, \sqrt{g} \, \Bigl( \epsilon^{\mu\nu\rho} 
\CA^i_{\mu} \, \partial_\nu \, c_\rho  -\i \, \CD^i  +\i \, \sigma^i \, H \Bigr) \,, \\
s_{\text{R-KK}} & \= \frac{1}{\pi^2} \, \int_{\CM_3}\dd^3 x \, \sqrt{g} \, 
\Bigl( \i \, v^\mu \, \bigl(\CA^{(R)}_\mu-\frac{1}{2}v_\mu \bigr))  
 -\frac{\i}{2} \, v^\mu \,v_\mu +\frac{\i}{2} \, H^2  -\frac{\i}{4} \, R^{(3)}  \Bigr) \,,\\
s_\text{KK-KK}& \= \frac{1}{\pi^2} \, \int_{\CM_3}\dd^3 x \, \sqrt{g}  \,  \Bigl( \i \, v^\mu \, c_\mu- 2 \, \i \, H \Bigr) 
\=  \frac{1}{\pi^2} \, \int_{\CM_3}\dd^3 x \, \sqrt{g}  \,  \Bigl( \epsilon^{\mu\nu\rho} \, 
c_\mu \,\partial_\nu \, c_\rho- 2\,\i \, H \Bigr) \,.
\label{eq:CSactionsKK}
\end{split}
\ee
The equations~\eqref{eq:CSactions1},~\eqref{eq:CSactions2},~\eqref{eq:CSactionsKK} together 
make up the complete list of CS terms.
As we explain below, when we have a KK reduction these actions can be combined 
together into a succinct expression in a natural manner. 

\vskip 0.4cm

The coefficients of the above actions are obtained by calculating the coefficients of the CS pieces,
which are obtained by integrating out all massive fermions that couple to the corresponding gauge fields. 
Integrating out a fermion $\tf$ with real-mass $m_\tf$ and charges $e^\tf_x, e^\tf_y$ under 
the gauge fields $\CA_x,\CA_y$ generates the term 
$\frac{1}{\pi^2} \, \int_{\mathcal{M}_3}\CA_x\wedge \mathrm{d}\CA_y$ 
with coefficient given by the one-loop exact formula 
(we follow the conventions of \cite{DiPietro:2014bca})
\begin{equation}
    -\frac{\i  \pi}{8} \, \sum_\tf \mathrm{sgn}(m_\tf) \, e^\tf_x \, e^\tf_y \,.
\label{eq:oneLoopCS}
\end{equation}
The contribution of the fermion to the coefficient of the gravitational CS term is given by 
(see Appendix~A of~\cite{Closset:2018ghr})
\begin{equation}
  -\frac{\i \,\pi}{192} \, \sum_\tf \mathrm{sgn}(m_\tf) \,. 
\label{eq:oneLoopCSgrav}
\end{equation}
The full effective action of the theory is the sum of the actions~\eqref{eq:CSactions1},
\eqref{eq:CSactions2} \eqref{eq:CSactionsKK} with 
coefficients obtained by summing~\eqref{eq:oneLoopCS}, \eqref{eq:oneLoopCSgrav} 
over all the massive fermions in the theory. 
(The actions with~$x$ and~$y$ different appear twice in the final action---as $x$-$y$ and~$y$-$x$---and 
therefore need to be multiplied by a factor of two.)

\vskip 0.4cm

The situation of interest in Section~\ref{sec:4dto3d} is the Kaluza-Klein reduction of a four-dimensional theory 
on a circle of radius~$R$. 
The bosonic fields in the three-dimensional vector multiplet are written 
in term of the 4d fields in~\eqref{eq:susy3dvec}, 
and the three-dimensional fermions are the reduction of the corresponding 4d fermions.
Consider a fermion of R-charge~$r_\tf$ transforming in a representation of weight~$\rho_\tf$ under the gauge group.
The tree-level real-mass of a KK mode of this fermion is given by (in the convention of~\cite{DiPietro:2014bca})
\be 
m_\tf \= -(
\rho_\tf\cdot A_Y - p_Y - r_\tf \, A^\text{nm}_Y -  \frac12 V^\text{nm}_Y)\,,
\label{realmass}
\ee 
where~$A^\text{nm}$, $V^\text{nm}$ are the 4d background R-gauge fields given in~\eqref{4dbackgnd}. 
Note that $p_Y$ also enters \eqref{eq:oneLoopCS} as the charge of the fermion 
under $U(1)_{\text{KK}}$.\footnote{
In the Euclidean context that we discuss here, the background fields~$A^\text{nm}$,~$V^\text{nm}$, and the 
effective radius~$R$ are complex. The definition~\eqref{realmass} is thought of as an analytic continuation,
and is read off from the coupling of the fermion to the other fields and parameters 
in the component of the covariant derivative along the KK direction~$Y$. The quantity~$m_\tf$
appears in the main text in formulas for the one-loop correction From the expression e.g.~\eqref{defSf},
we see that it appears only as the function~$\sgn(m_\tf)$. Then the 
the~$\sgn$ function needs to be appropriately defined. As discussed below Equation~\eqref{eq:KKsumBbar},
we can do that by defining it for real~$R$, and then continuing the formulas to complex~$R$.} 
Since we take~$R \to 0$ at the end of the calculations, it is enough to keep only the singular 
pieces in the formula~\eqref{realmass}.

Using these relations we proceed to write the three-dimensional effective action directly in terms of the 
dynamical 4d fields. 
The contribution of the actions coming from~\eqref{eq:CSactions1} and~\eqref{eq:CSactionsKK} 
to the full effective action can be written as the sum of the following two actions,
\be
\begin{split}
\wt{S}^{\,\tf}_\text{g-g} 
	& \= -\i \pi \, \frac{\mathrm{sgn}(m_\tf)}{8}\Bigl(\rho_\tf^i \, \rho_\tf^j \, s^{ij}_\text{g-g} 
		+ 2 \, p_Y \, \rho_\tf^i \, s^i_\text{g-KK} + p_Y^2 \, s_\text{KK-KK} \Bigr) \\
	&\=-\i \pi \, \frac{\mathrm{sgn}(m_\tf)}{8}\biggl( \bigl(\rho_{\tf}\cdot A_Y - p_Y\bigr)^2
		\int_{\mathcal{M}_3}\dd^3 x \,\sqrt{g}\,\bigl(\i \,v^\mu   c_\mu -2\i \, H\bigr)\\
	&\kern6em~~~~+2 \bigl(\rho_{\tf}\cdot A_Y - p_Y\bigr)\int_{\mathcal{M}_3}\dd^3 x \,\sqrt{g}\, 
		\bigl(-\i \, v^\mu \, (\rho_{\tf}\cdot A_\mu) +\i \,(\rho_{\tf}\cdot D)\bigr)\\
	&\kern6em~~~~+\int_{\mathcal{M}_3}\dd^3 x \,\sqrt{g}\, 
		\bigl(\epsilon^{\mu\nu\rho}(\rho_{\tf}\cdot A_\mu)\, \partial_\nu \, (\rho_{\tf}\cdot A_\rho)\biggr) \,,
\label{S:susy:explicit1}%
\end{split}
\ee
\be
\begin{split}
2 \, \wt{S}^{\,\tf}_\text{g-R} 
	& \=  -2 \, \i \pi \, \frac{\mathrm{sgn}(m_\tf)}{8} \, r_\tf \, \Bigl(\rho_\tf^i \, s^{i}_\text{g-R} 
			+ p_Y \, s_\text{R-KK}  \Bigr) \\
	&\=- 2 \, \i \pi \, \frac{\mathrm{sgn}(m_\tf)}{8} \, r_\tf\\
	&\kern2em~~\biggl( \bigl(\rho_{\tf}\cdot A_Y - p_Y\bigr)
	\int_{\mathcal{M}_3}\dd^3 x \,\sqrt{g}\,
	\Bigl(-\i \,v^\mu\big(\mathcal A^{(R)}_\mu-\frac{1}{2}v_\mu\big)
	+ \i \,\frac{1}{2}v^\mu \, v_\mu - \i \, \fft12H^2 + \i \,\fft14R^{(3)} \Bigr)\\
	&\kern6em~~+\int_{\mathcal{M}_3}\dd^3 x \,\sqrt{g}\, \Bigl(\epsilon^{\mu\nu\rho}(\rho_{\tf}\cdot A_\mu) \,
	\partial_\nu \, \bigl(\mathcal A^{(R)}_\rho-\fft12v_\rho\bigr)+\i \, (\rho_{\tf}\cdot D)H \Bigr)\biggr) \,.
\label{S:susy:explicit2}%
\end{split}
\ee

Finally we specialize to the BPS configurations considered in the main text, given 
in~\eqref{AYval3d}. The above two terms take the following value 
\be
\begin{split}
\wt{S}^\tf_\text{g-g} & \=-\i \pi \, \frac{ \mathrm{sgn}(m_\tf)}{8R^2} \,  
\bigl(\rho_{\tf}\cdot \uu - k_Y \bigr)^2\, A_{\mathcal{M}_3} \,, \\
2 \, \wt{S}^\tf_\text{g-R} & \=-\i \pi \,  \frac{\mathrm{sgn}(m_\tf)}{8R} \, 
2 \, r_\tf  \, \bigl(\rho_{\tf}\cdot \uu - k_Y \bigr) \; L_{\mathcal{M}_3} \,,
\end{split}
\label{S1S2fin}
\ee
where~$A_{\mathcal{M}_3}$ and $L_{\mathcal{M}_3}$ are functions
of the three-dimensional background\footnote{To compare with~\cite{DiPietro:2016ond} 
note that~$A_{\mathcal{M}_3}=- A_{\mathcal{M}_3}^{\text{there}}$ and 
$L_{\mathcal{M}_3}=\i\, L_{\mathcal{M}_3}^{\text{there}}.$},
\begin{equation}
\label{defALl}
    \begin{split}
    A_{\mathcal{M}_3}& \= \frac{1}{\pi^2}\int_{\mathcal{M}_3} \dd^3x \, \sqrt{g} \, \bigl(\i \,v^\mu \, c_\mu - 
    2 \, \i \, H\bigr)\,, \\
    L_{\mathcal{M}_3}& \= \frac{1}{\pi^2}\int_{\mathcal{M}_3} \dd^3x \, \sqrt{g} \,
    \Bigl(-\i \, v^\mu\mathcal A^{(R)}_\mu + \i\, v^\mu \, v_\mu - \i \, \fft12H^2 + \i \, \fft14R^{(3)} \Bigr) \,.
    \end{split}
\end{equation}

\vskip 0.4cm

We now turn to the remaining terms in the full action, namely those coming from 
the terms in~\eqref{eq:CSactions2},
\be
\begin{split}
    S^\tf_\text{R-R}&\=-\i \pi \, \frac{\mathrm{sgn}(m_\tf)}{8} \, \bigl(r_\tf^2-\frac{1}{6}\bigr) \, R_{\CM_3}  \,, \\
  S^\tf_\text{grav}&\=-\i \pi \, \frac{\mathrm{sgn}(m_\tf)}{192} \, G_{\CM_3} \,.
\label{S:susy:explicit3}%
\end{split}
\ee
Note that  both~$R_{\CM_3}$ and~$G_{\CM_3}$ contain~$\CA^{(R)}\wedge\mathrm{d} \CA^{(R)}$ terms, 
and it is the sum of the corresponding coefficients that is fixed by~\eqref{eq:oneLoopCS}. 
Since the coefficient in~$S^\tf_{\text{grav}}$ is fixed by~\eqref{eq:oneLoopCSgrav}, the shift~$-1/6$ 
in the coefficient of~$S^\tf_{\text{R-R}}$ serves to cancel the~$\CA^{(R)}\wedge\mathrm{d} \CA^{(R)}$ 
term coming from~$S^\tf_{\text{grav}}$.
The final result for the action of the BPS configurations 
up to~$\rO(R^0)$ obtained by integrating out a fermion~$\tf$ is
given by the sum of the actions in~\eqref{S1S2fin},~\eqref{S:susy:explicit3}.

\section{Values of supersymmetrized Chern-Simons actions \label{sec:CSactionvals}}

In this appendix we record the values of various terms in the supersymmetrized actions 
of Appendix~\ref{sec:CSactions} evaluated on the configurations discussed in 
Section~\ref{sec:4dto3d}. 
We first recall from \S\ref{sec:4dto3d} the values of the various fields entering the actions. 
The three-dimensional metric is 
\be 
\dd s_3^2  \=  \dd \theta^2 + \sin^2 \theta \,  \dd \phi_1^2 + \cos^2 \theta \, \dd \phi_2^2  - c^2 \,,
\ee
the graviphoton and its Hodge dual are
\be  
c \= 
 -\i \frac{\O}{\sqrt{1-\O^2}} \, \bigl( \sin^2 \theta \,\dd \phi_1  + \cos^2 \theta \,\dd \phi_2 \bigr) \,,
\qquad 
 v \= \frac{2 \, \i}{\sqrt{1-\O^2}} \, c \,. 
\ee
The auxiliary background supergravity multiplet fields are  
\be 
H \=  -\frac{\i}{\sqrt{1-\O^2}} \,, \qquad 
\CA^R_\mu \=  \Bigl(\frac{\t}{R} - \frac{n_0}{2R}  + \frac{\i}{\sqrt{1-\O^2}} \Bigr)  \, c_\mu \,.\label{eq:HandARmuAppC}
\ee
The four-dimensional gauge fields are 
\be 
A^i_Y \= \frac{u_i}{R} \,, \qquad D^i \= 0 \,, 
\ee

The Chern-Simons action for~$c$  
\be
S^\text{CS} (c)  \= \int_{\CM_3} \, c \wedge \dd c  \= \i \int_{\CM_3} \dd^3 x \,  \sqrt{g} \, v^\mu \, c_\mu\,,
\ee
evaluates to 
\be \label{CScval}
 \frac{1}{4 \pi^2} \, S^\text{CS} (c) \= 
 \=  \frac{ \O^2}{1-\O^2}  \= -1 + \rO(\rr)  \,.
\ee
The other building blocks for the actions of the background fields in the three-dimensional 
theory are given below, including their limiting behavior as~$\rr \to 0$ with~$\t$ fixed (i.e.~as~$\O \to \infty$),
\be
\frac{1}{4 \pi^2} \, S^{(H)} \= \frac{\i}{4\pi^2} \int_{\CM_3} \dd^3 x \,  \sqrt{g} \, H \= \frac{1}{2(1-\O^2)} 
 \=  \rO(\rr) \,,
\ee
\be
\frac{1}{4 \pi^2} \, S^{(v)} \= \frac{1}{4\pi^2}\int_{\CM_3} \dd^3 x \,  \sqrt{g} \, v^\mu \, v_\mu \=  \frac{ \O^2}{(1-\O^2)^{\frac32}} 
 \=  \rO(\rr) \,,
\ee
\be
\frac{1}{4 \pi^2} \, S^{(H^2)} \= \frac{1}{4\pi^2}\int_{\CM_3} \dd^3 x \,  \sqrt{g} \, H^2 \= -\frac{1}{2(1-\O^2)^\frac32} 
 \=  \rO(\rr) \,,
\ee
\be
\frac{1}{4 \pi^2} \, S^{(R)} \= \frac{1}{4 \pi^2}\int_{\CM_3} \dd^3 x \,  \sqrt{g} \, R^{(3)} \= -\frac{-6 + 8 \O^2}{2(1 - \O^2)^\frac32} 
 \=  \rO(\rr) \,.
\ee

\section{Dimensional reduction for the case~$\Omega_1 \neq \Omega_2$ \label{sec:diffomegas}}

We begin by writing the background configuration in~\eqref{4dbackgnd} as a KK compactification 
to three dimensions, i.e.~a circle fibration on a 3-manifold~$M_3$. 
We have
\be \label{4dKK12}
\dd  s_4^2  \=  \dd s_3^2 + \rme^{2 \phi} (\dd t_E + \wt c \,)^2 \,,  
\ee 
where the metric on~$M_3$ is
\be \label{3dmetric12}
\begin{split}
\dd s_3^2  \= \wt g_{\mu\nu} \, \dd x^\mu \, \dd x^\nu 
\=  \dd \theta^2 + \sin^2 \theta \,  \dd \phi_1^2 + \cos^2 \theta \, \dd \phi_2^2  - \rme^{2\phi} \, \wt c^{\,2} \,,\\
\end{split}
\ee
and the graviphoton and KK scalar are
\be \label{3dKKbackgnd12}
\begin{split}
\rme^{2\phi} & \= 1 - \Omega_1^2  \, \sin^2 \theta - \Omega_2^2  \, \cos^2 \theta \,, \\
\wt c \= \wt c_\mu \, \dd x^\mu 
& \= -\i\, \rme^{-2\phi} \bigl( \Omega_1 \,\sin^2 \theta \,\dd \phi_1  + \Omega_2 \,\cos^2 \theta \,\dd \phi_2 \bigr) \,.
\end{split}
\ee
For the case~$\Omega_1=\Omega_2 = \O$, we have that~$e^{2\phi} = 1-\O^2$, so that the graviphoton~$c$ 
defined in~\eqref{4dmetricKKform} is related to~$\wt c$ as~$c=\rme^\phi \, \wt c$. 
The magnitude of the volume form in three dimensions is 
\be \label{sqrtg12}
\sqrt{\wt g} \= \frac12 \, \rme^{-\phi} \, \sin 2 \theta  \,.
\ee

The associated Chern-Simons action 
\be
S^\text{CS} ( \, \wt c \,)  \= \int_{\CM_3} \, \wt c \wedge \dd \wt c  
\= \i \int_{\CM_3} \dd^3 x \,  \sqrt{\wt g} \; \wt v^\mu \, \wt c_\mu 
\ee
(where~$\wt v = - \i * \dd \wt c $ is the Hodge dual) evaluates to 
\be \label{CScval12}
S^\text{CS} (\, \wt c \,) \= 4 \pi^2 \,  \O_1 \, \O_2 \int_0^{\pi/2} 
\frac{\sin 2 \theta}{(1-\O_1^2 \sin^2 \theta -\O_2^2 \cos^2 \theta)^2} \, \dd \theta  
\= 4 \pi^2 \frac{\O_1 \O_2}{(1-\O_1^2)(1-\O_2^2)}  \,.
\ee

For the identification between the four-dimensional and the three-dimensional fields, we follow the treatment of~\cite{Assel:2014paa} 
applied to the metric~\eqref{4dKK12}.
The result is
\be
\frac12 \, \rme^\phi \, \wt v_\mu \= V^\text{nm}_\mu - V^\text{nm}_{t_E} \, \wt c_\mu \,, \qquad 
\wt H \= \rme^{-\phi} \,V_{t_E} \,, \qquad 
\CA^R_\mu \= A^\text{nm}_\mu -  A^\text{nm}_{t_E}   \; \wt c_\mu  + \frac12 \, \rme^\phi \, \wt v_\mu \,.
\ee
The values of these fields are
\be \label{vHvalues12}
 \rme^\phi \, \wt v_\mu \=  - 2 \, V^\text{nm}_Y \, c_\mu  \=  2\,\i \, \wt c_\mu \,, \qquad  
\wt H \= - \i \, e^{-\phi} \,,
\ee
\be \label{ARval12}
 \CA^R_\mu \=   \i  \bigl( \frac12 (\O_1 +\O_2) - 1 \bigr) \, \wt c_\mu   + \frac12 \, \rme^\phi \, \wt v_\mu 
 \=   \frac{\i}{2} (\O_1 +\O_2) \, \wt c_\mu \,.
\ee
We can now calculate the various actions as in Appendix~\ref{sec:CSactionvals}, and we 
find that, in the~$\rr \to 0$, $\Omega\to\infty$ limit we have the effective replacement 
\be
\frac{1}{R^2} \to \frac{1}{\t \, \s} \,, \qquad \frac{1}{R} \to \frac{\t + \s}{2\t \, \s}  
\ee
in the effective potential~\eqref{Veffans3d}.

\bibliographystyle{JHEP}

\providecommand{\href}[2]{#2}\begingroup\raggedright\endgroup

\end{document}